# The conformal manifold of three dimensional $\mathcal{N} = 4$ supersymmetric star-shaped-quiver theories

Tal Miller

# The conformal manifold of three dimensional $\mathcal{N}=4$ supersymmetric star-shaped-quiver theories

Research Thesis

Submitted in partial fulfillment of the requirements
for the degree of Master of Science in Physics

Tal Miller

Submitted to the Senate
of the Technion — Israel Institute of Technology
Heshvan 5780     Haifa     November 2019

# Contents

# List of Figures

# List of Tables

# Abstract

In this thesis we calculate the dimension of the conformal manifold (DCM) for a class of 3d $\mathcal{N} = 4$ supersymmetric theories called the star-shaped-quiver (SSQ) theories. These theories are the mirror dual theories of class $\mathfrak{s}$ theories, constructed as a compactification of the 4d class $\mathcal{S}$ theories on a circle $\mathbb{S}^1$. The 4d class $\mathcal{S}$ theories themselves are constructed as a compactification of the 6d $\mathcal{N} = (2,0)$ superconformal field theories on a Riemann surface $C_{g,s}$ (with genus $g$ and $s$ punctures). The IR fixed point of these theories can be strongly coupled, and an interesting probe of the fixed point is the conformal manifold, the space of all exactly marginal deformations. The supersymmetric index is a tool, developed in recent years, that allows to calculate the DCM for supersymmetric theories. The index encodes within it the exactly marginal operators, but to deduce the DCM involves a few more tools that transform the problem into a group-theoretic one. We employ these tools for the SSQ theories, first calculate the supersymmetric index and then calculate the DCM. Our results are that the DCM of the 3d SSQ (or class $\mathfrak{s}$) theories scales as $\sim g^4$ and $\sim s^2$, which is significanly larger that the DCM of the related 4d class $\mathcal{S}$ theories which scales linearly with $g$ and $s$.

# Abbreviations and Notations

| | | |
|---|---|---|
| QFT | : | quantum field theory |
| CFT | : | conformal field theory |
| SCFT | : | superconformal field theory |
| CM | : | conformal manifold |
| DCM | : | dimension of the conformal manifold |
| SSQ | : | star-shaped-quiver |
| PE | : | Plethystic exponent |

# Chapter 1

# Introduction

Quantum field theories (QFTs) exhibit various behaviors in the low energy limit, like confinement and strong coupling, which are difficult to calculate [1]. Supersymmetry (also called super-Poincaré symmetry) is the unique extension of the Poincaré symmetry in more than 2 dimensions [2]. Supersymmetric QFTs display similar phenomena to non-supersymmetric QFTs, but due to the enhanced symmetry they can serve as a "theoretical laboratory" for strongly-interacting QFTs [3]. Calculating non-trivial properties of strongly-interacting theories is one aim of the theoretical high-energy and condensed matter physics communities, and at least a subset of such properties can sometimes be calculated exactly for supersymmetric QFTs.

The low/high energy (IR/UV) fixed points of QFTs are scale-invariant theories, which are usually symmetry-enhanced to conformal field theories (CFTs) [4][5][6]. For supersymmetric theories, the fixed point theories are symmetry-enhanced to superconformal field theories (SCFTs) [7], the result of the combination of supersymmetry and conformal symmetry.

Given a CFT, the different couplings that can be added to the theory (terms in the Lagrangian for Lagrangian CFTs), while keeping it conformal, spans a space of theories called the conformal manifold (CM) $\mathcal{M}_c$ [8]. This manifold is spanned by the exactly marginal couplings (irrelevant or marginally-irrelevant couplings give the same theory, and the relevant or marginally-relevant couplings break the conformal symmetry). The dimension of the conformal manifold (DCM) is the number of independent exactly marginal couplings. In other words, the DCM is the number of tunable parameters the CFT has. The number of these parameters and the structure of the conformal manifold are an important set of parameters defining the CFT. Sometimes these parameters can be understood by thinking of the theory as a compactification of a higher dimensional theory, and then they can be related to geometric properties of the setup (e.g. [9]).

The highest dimension where interacting SCFTs can be defined is 6 dimensions [10][7]. We begin with the 6d $\mathcal{N} = (2,0)$ SCFT, which has maximal supersymmetry (16 supercharges) without spin$> \frac{1}{2}$ multiplets. This theory is actually not unique but has an ADE algebra classification, and we only consider type $A$ theories in this thesis.

These theories do not have a known Lagrangian description [11]. Compactifying on a torus preserves all the supersymmetry and gives a 4d $\mathcal{N} = 4$ theory (since the torus is a flat manifold, translation symmetry remains unbroken in all directions, and therefore the momentum and supersymmetry generators are unbroken). Compactifications on a general punctured Riemann surface $C_{g,s}$ (with genus $g$ and $s$ punctures) breaks some of the supersymmetry and gives 4d $\mathcal{N} = 2$ theories, called class $\mathcal{S}$ theories [12]. Most of these theories do not have a known Lagrangian description. In this work we will be interested in the further compactification of theories of class $\mathcal{S}$ on a circle $\mathbb{S}^1$ to 3d $\mathcal{N} = 4$ theories, called class $\mathfrak{s}$ theories or Sicilian theories (which are also generally non-conformal). These theories are conjectured to be mirror dual to the 3d $\mathcal{N} = 4$ star-shaped-quiver (SSQ) theories [13][11]. This means they share the same IR fixed point CFT and therefore the same CM. SSQ theories do have a Lagrangian description, and therefore it will be easier to analyze their IR properties [11]. The aim of this work will be to calculate the DCM of the SSQ theories.

A diagramatic description for the relations between the theories we discussed and their conformal manifolds:

Figure 1.1: Relations between different theories and their conformal manifolds

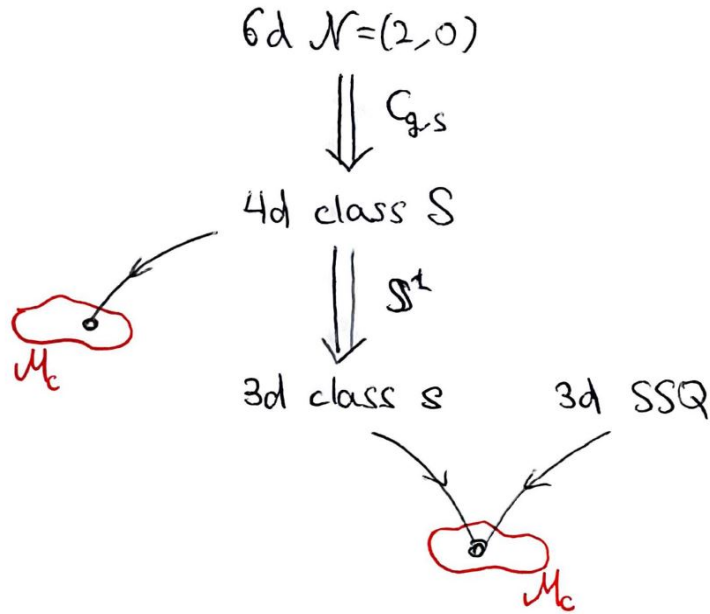

In the diagram, double lines (and the labels next to them) represent a compactification of a theory, and single lines represent RG-flows to IR fixed points (which reside in the CM of the theory).

The probe we shall use to study the IR strong coupling regime is the supersymmetric index, which is invariant under renormalization group (RG) flow and therefore can be calculated in the weakly coupled UV for such theories (in 3d all gauge theories are UV free, unlike the 4d case where it depends on the gauge group and matter content) [14]. In essence, the supersymmetric index is the partition function of the theory, calculated on a compact manifold, rather than on flat space [15][16][11]. The index is a non-trivial function of the theory's properties (fields, couplings and symmetries) and among other properties it encodes the marginal couplings and the global symmetry group in the IR [17][18], as discussed in chapter 3. These properties will allow us to calculate the DCM.

Generally in QFTs, marginal operators can be either marginally relevant, marginally irrelevant or exactly marginal. Considering only supersymmetric operators, in [8] it is proven using conformal perturbation theory that in SCFTs in 2,3,4 dimensions (with at least 4 supercharges) there are no marginally relevant operators at all, only marginally irrelevant or exactly marginal operators exist (above 4 dimensions there are no exactly marginal couplings at all [19]). Moreover, it is shown in [8] that considering the quotient of the space of marginal couplings of the SCFT $\{\lambda\}$ by the complexified global symmetry group $G^{\mathbb{C}}$ is locally equivalent to the space of exactly marginal couplings, which is precisely the conformal manifold $\mathcal{M}_c = \{\lambda\}/G^{\mathbb{C}}$. The dimensions of both spaces are the same.

Calculating the CM now becomes a group theoretic problem, given we know the global symmetry. For a UV-free QFT that has a Lagrangian, the global symmetry can be deduced directly from the Lagrangian. Flowing to the strongly-coupled IR the global symmetry may even be enhanced [20][21]. Therefore, it is not trivial to find the global symmetry in the IR of the SSQ theories. Luckily, the supersymmetric index encodes the global symmetry group in the IR as well (up to some caveats), as we explain in chapter 3. Finally, to extract the DCM we will use the Hilbert series, which performs the above mentioned quotient and encodes the DCM, as we explain in chapter 4. The IR fixed point of a SSQ theory is a point on the CM, that has at least as much global symmetry as in the UV. Moving to different points (theories) on the CM, means turning on couplings that might break some or all of the global symmetry. If the global symmetry is indeed completely broken when moving to a generic point on the CM, then we expect the DCM to equal the number of marginal couplings minus the dimension of the global symmetry group. The result we obtain from our analysis is consistent with the global symmetries being completely broken on a generic point on the CM of the SSQ theories (happens due to the large representations involved).

We would also like to understand how the DCM can be understood from a geometric perspective. The DCM of the 4d class $\mathcal{S}$ theories was interpreted as a combination of some geometric invariants of the Riemann surface $C_{g,s}$ the theory is defined with [9]. Similarly, we imagine the DCM of the 3d class $\mathfrak{s}$ theories can be interpreted using the geometric invariants of the Riemann surface times a circle $C_{g,s} \times \mathbb{S}^1$ the theory is defined with, but such an interpretation has yet to reveal itself to us.

The structure of the thesis is as follows: In chapter 2 we define supersymmetry and superconformal symmetry which are essential to the SSQ theories. In chapter 3 we define the supersymmetric index, discuss its properties, calculate it for the SSQ theories and conjecture the form of the index for the general case. In chapter 4 we define the Hilbert series and calculate the DCM of the SSQ theories. Finally, in chapter 5 we summarize the work.

# Chapter 2

# Supersymmetry

In this thesis we mainly deal with supersymmetric and superconformal 3d theories, so we define these symmetries schematically before we continue.

## 2.1 Supersymmetry algebra

In supersymmetry we add spinor generators $Q$ (also called supersymmetry generators) to the Lorentz algebra (or the Poincaré symmetry when we also consider translations), forming the super-Poincaré algebra. These generators satisfy non-trivial anti-commutation relations that connect them to the momentum generator $P$ in the schematic relation $\{Q, Q\} \propto P$ [2]. Non-formally, the spinor generators are the "square roots" of the momentum generator.

More concretely, in 3d the spinor has 2 real components, so we define the spinor generator $Q_\alpha$ (which holds 2 supercharges) whose commutation relations are [22]

$$\{Q_\alpha, Q_\beta\} = 2\gamma^\mu_{\alpha\beta} P_\mu$$
$$[Q_\alpha, P_\mu] = 0$$
(2.1)

where $P_\mu$ are the Lorentz momentum generators and $\gamma^\mu$ are the Clifford matrices. These matrices satisfy the Clifford algebra $\{\gamma^\mu, \gamma^\nu\} = 2\eta^{\mu\nu}$ (where $\eta^{\mu\nu}$ is the Minkowski spacetime metric) and in 3d they are simply the $2 \times 2$ Pauli matrices, unlike the $4 \times 4$ matrices in 4d. The algebra we wrote is the minimal supersymmetry in 3d, meaning we have $\mathcal{N} = 1$ multiples of the minimal amount of supercharges. In this case there is no R-symmetry and the number of complex supercharges is not enough to use holomorphy properties [23] and the results of [8] for the CM (mentioned in the introduction).

Defining supersymmetry with $\mathcal{N} > 1$ multiples of the minimal amount of supercharges $N_Q$ is called extended supersymmetry (true in any number of dimensions). In 3d $\mathcal{N} = 2$ we define a complex spinor generator $Q_\alpha$ such that it and its conjugate $\tilde{Q}_\alpha$ are the new generators (4 supercharges in total), whose commutation relations are

[24][25]

$$\begin{aligned} \left\{Q_\alpha, \tilde{Q}_\beta\right\} &= 2\gamma^\mu_{\alpha\beta}P_\mu + 2i\epsilon_{\alpha\beta}Z \\ \{Q_\alpha, Q_\beta\} &= \left\{\tilde{Q}_\alpha, \tilde{Q}_\beta\right\} = 0 \\ [Q_\alpha, P_\mu] &= \left[\tilde{Q}_\alpha, P_\mu\right] = 0 \end{aligned} \quad (2.2)$$

where $\epsilon_{\alpha\beta}$ is the anti-symmetric tensor and $Z$ is a central charge. There is now an additional global symmetry which is the $U(1)$ R-symmetry with a generator $R$ whose property is that it does not commute trivially with the spinor generator:

$$[R, Q_\alpha] = -Q_\alpha \quad (2.3)$$

Higher $\mathcal{N}$ algebras are also possible, and we will be interested in 3d $\mathcal{N}=4$ where the number of supercharges is doubled and there are more central charges $Z^{ij}$:

$$\left\{Q^i_\alpha, \tilde{Q}^j_\beta\right\} = 2\delta^{ij}\gamma^\mu_{\alpha\beta}P_\mu + 2i\epsilon_{\alpha\beta}Z^{ij} \quad (2.4)$$

The R-symmetry is enhanced to $SO(4) \sim SU(2) \times SU(2)$, meaning there are two R-symmetry generators that form the algebra. The SSQ theories we deal with in this thesis are $\mathcal{N}=4$.

Generally, $\mathcal{N}$ is correlated with the size of the R-symmetry group, and what that group is depends on the dimension $d$ [10][26]. The number of supercharges is the rank of the R-symmetry group, and the supercharges always transform in the fundamental representation of the R-symmetry group.

## 2.2 Superconformal algebra

Conformal symmetry for $d$ spacetime dimensions is (locally isomorphic to) $SO(d,2)$, which has rank$(SO(d+2)) = \lfloor (d+2)/2 \rfloor = \lfloor d/2 \rfloor + 1$ Cartan generators. It is common to divide the generators to the dilatation generator $\Delta$ and the rotation generators $M_{\mu\nu}$ of $SO(d-1,1)$ (normal spacetime), their number being rank$(SO(d)) = \lfloor d/2 \rfloor$ (the generators with indices $\mu,\nu$ transform in the antisymmetric second rank tensor representation of $SO(d-1,1)$).

Additional generators in the conformal algebra (that are not independent) are the momentum generators $P_\mu$ and the special conformal generators $K_\mu$. The commutation relations of all these generators are non-trivial and we refer the reader to [7] for the full list of these relations.

Combining supersymmetry with conformal symmetry, means the supercharges need to be defined as spinor representations of (the double cover group of) $SO(d,2)$ rather than of $SO(d-1,1)$, which effectively means the number of supercharges is doubled compared to the spinors of $SO(d-1,1)$ [26][27]. In the formalism of $SO(d-1,1)$, the additional supercharges $S_{i\alpha}$ are called special supersymmetry generators, and their number is the same as the $Q_{i\alpha}$ generators.

In the superconformal algebra $Q$ and $S$ have non-trivial commutation relations with each other and with the rest of the conformal generators. The exact forms of these relations depend on the spacetime dimension $d$, and we refer the reader to [7] for the full list of these relations. The schematic form relevant for all dimensions is $\{Q,Q\} \propto P$ (same as in normal supersymmetry algebra), $\{S,S\} \propto K$, and $\{Q,S\}$ is proportional to a combination of generators including $M$, $\Delta$ and the R-symmetry generators [7].

Superconformal algebras for interacting theories (with spin $\leq 1$ fields) exist in dimensions $d \leq 6$. For $d = 2$ the Lie-algebra is infinite-dimensional (super-Virasoro algebra [28]) which is a different subject we do not deal with in this thesis. For different dimensions $d = 3, 4, 5, 6$ we summarize the number of independent supercharges $N_Q$ for minimal supersymmetry $\mathcal{N} = 1$, and the R-symmetry group for general $\mathcal{N}$ [7][29][30]:

Table 2.1: R-symmetry groups for different dimensions and numbers of supercharges in superconformal algebras

| $d$ | $N_Q$ | R-symmetry group | rank |
|---|---|---|---|
| 3 | 2 | $SO(\mathcal{N})$ | $\lfloor \mathcal{N}/2 \rfloor$ |
| 4 | 4 | $\begin{cases} SU(\mathcal{N}) \times U(1) & \mathcal{N} \neq 4 \\ SU(4) & \mathcal{N} = 4 \end{cases}$ | $\begin{cases} \mathcal{N} & \mathcal{N} \neq 4 \\ 3 & \mathcal{N} = 4 \end{cases}$ |
| 5 | 8 | $\begin{cases} Sp(1) = SU(2) & \mathcal{N} = 1 \\ \text{none exist} & \mathcal{N} > 1 \end{cases}$ | $\begin{cases} 1 & \mathcal{N} = 1 \\ \text{none exist} & \mathcal{N} > 1 \end{cases}$ |
| 6 | 8 | $Sp(2\mathcal{N})$ | $\mathcal{N}$ |

Specific popular cases in 3,4d are:

- For 4d, the $\mathcal{N} = 1$ theory has $U(1)$ R-symmetry, $\mathcal{N} = 2$ has $SU(2) \times U(1)$, and $\mathcal{N} = 4$ has $SU(4)$.

- For 3d, $\mathcal{N} = 2$ has $SO(2) \sim U(1)$ R-symmetry (same amount of supersymmetry as in 4d $\mathcal{N} = 1$), $\mathcal{N} = 3$ has $SO(3) \sim SU(2)$, and $\mathcal{N} = 4$ has $SO(4) \sim SU(2) \times SU(2)$ (same amount of supersymmetry as in 4d $\mathcal{N} = 2$).

## Chapter 3

# Supersymmetric Index

## 3.1 Introduction

Our goal is to calculate the DCM for a class of theories called the SSQ theories. To that end, we shall define and calculate the supersymmetric index which we show encodes the DCM. In essence, the supersymmetric index can be defined for supersymmetric QFTs in two ways:

1. As the partition function of the theory, calculated on a compact manifold rather than on flat space, using supersymmetric localization technique. For dimension $d$ the compact manifold is $\mathbb{S}^{d-1} \times \mathbb{S}^1$ [31][15][16].

2. As a trace over the Hilbert space (summing states) of the theory weighed by fugacities which correspond to some symmetries. When calculating this in the IR, it is called the superconformal index. However, if the theory is strongly coupled in the IR then we do not necessarily know the Hilbert space. Here, the fact that the index is invariant under RG flow comes to the rescue, and allows to perform the calculation in the UV where the fields are weakly coupled [14]. This works provided the R-symmetry is chosen such that the IR theory is indeed superconformal [32][33].

Depending on the dimension, the calculation of the index is easier with different approaches:

- Using the trace formula seems easier because the calculation involves counting the possible local operators/states, rather than using the partition function which involves calculating a path integral. This is indeed the case in 4d [15].

- However, in 3d and 5d there are also non-trivial operators/field configurations (monopoles and instantons, respectively) that contribute to the index. In these dimensions the index was first calculated using the partition function and later interpreted as a trace formula, meaning a sum of contributing states where some of the states are the non-trivial ones [34][16][35].

- In 6d there are no known interacting theories with a Lagrangian description and so the index can be inferred only indirectly [36].

- In 2d, the index was calclated using the partition function, but (to date) the result did not allow to be interpreted as a trace formula at all [31].

The definition of the superconformal index (the second way to define the supersymmetric index) is a trace over the Hilbert space

$$I \equiv \text{Tr}_{\mathcal{H}} \left( (-1)^F z^\delta \prod_i a_i^{h_i} \right) \quad (3.1)$$

where $F$ is the fermion number, $\delta$ the charge under the operator $\{Q, Q^\dagger\}$, $\{h_i\}$ are charges under spacetime related operators that commute with $\{Q, Q^\dagger\}$ and $z, \{a_i\}$ are fugacities (parameters). In 3d $\mathcal{N} = 2$ there is only one operator that commutes with $\{Q, Q^\dagger\}$ (a linear combination of Lorentz generators and the $U(1)$ R-symmetry generator) whose fugacity we denote by $q$. If the theory has additional global symmetries their fugacities can also be added to the definition of the index in the same way. For a more elaborate discussion on the definition of the superconformal index see Appendix B.

The usefulness of the index to calculate the DCM stems from the fact that in the superconformal IR the only superconformal multiplets that contribute to the index in power $q^1$ (in the power expansion of the index in $q$) are the marginal operators with a positive sign, and the conserved currents (of the global symmetry) with negative sign [17][18]. As we further discuss in chapter 4, we require to know both the marginal operators and the global symmetry group $G$ to calculate the DCM.

As discussed in the introduction (chapter 1), if the theory is strongly coupled in the IR, the global symmetry identified in the free UV might be enhanced in the IR by an a priori unknown amount [20][21]. A subset of the fugacities the index depends on are the maximal torus variables of the global symmetry group $G$ (see Appendix A.1). If the global symmetry is identified correctly, the maximal torus variables should form characters of representations of $G$. This is a first consistency check both for the integrity of the index and for the global symmetry group $G$ being correctly identified. The negative contribution in power $q$ are the conserved currents, so it should appear as a character of the adjoint representation of $G$. This is a second consistency check for the correct identification of $G$. Overall, the contribution to the supersymmetric index in first power in $q$ should take the form [17][18]:

$$I_q = \sum_i \chi_{R_i(G)}^{\text{marg}} - \chi_{\text{adj}(G)} \quad (3.2)$$

From this we can extract the marginal operators and their representations $\{R_i\}$. In chapter 4 we show how to calculate the DCM from $I_q$. A caveat to the process is that if the symmetry is enhanced in the IR we might not notice it, which could change the DCM. We discuss the effect of this in Appendix C.1, but everywhere else assume that the symmetry does not enhance in the IR. Another possibility is that in the IR the symmetry (global symmetry or even supersymmetry) is decreased rather than enhanced, which could happen due to spontaneous symmetry breaking in the vacuum state. Since the supersymmetric index essentially counts vacuum states, symmetry breaking would generically be detectable in a messed up structure of the index. This does not appear to happen in the theories discussed in the thesis.

R-symmetry can sometimes mix with $U(1)$ global symmetries. As we mentioned, we can calculate the supersymmetric index using the weakly coupled UV fields, given we pick the R-charges at the values that will make the theory superconformal in the IR. The method of $Z$-extremization can calculate the required R-charges [32]. However, for 3d theories with a non-abelian R-symmetry we do not expect a mixing to be possible at all, since non-abelian global symmetries cannot mix. The theories we will consider in this thesis (defined further down this chapter) all have non-abelian R-symmetry and so we do not expect any issues. Nevertheless, we will validate this for a few of the theories we consider, using the $Z$-extremization procedure.

The content of the rest of this chapter is as following:

- In section 3.2, we review the prescription that allows to calculate the index for any 3d theory, using the basic building blocks of the index.

- In section 3.3, we define the $T[SU(N)]$ theories, which are one of the building blocks of the SSQ theories. We discuss their global symmetry and calculate their index.

- In section 3.4, we define the SSQ theories, discuss their global symmetry, calculate their index and conjecture a general form.

- In section 3.5, we review the method of $Z$-extremization and calculate the superconformal R-charges for several theories.

## 3.2 Building blocks of the 3d index

We begin with a brief exposition on the building blocks of QFTs in general. Not all QFTs have a known Lagrangian description [11], however, Lagrangian QFTs are more intuitive and there are well defined building blocks we can use when constructing them. Lagrangian QFTs are defined using the field content and the Lagrangian. The latter can either be chosen explicitly, in which case the global symmetries emerge, or global symmetries can be imposed, which restricts the possible terms in the Lagrangian. An additional secondary choice is the background metric (i.e. Euclidean or Minkowski signatures). Requiring the theory to be quantum-mechanically well-defined at all energy scales (renormalizable) further constrains the possible terms in the Lagrangian (only the relevant operators) [1]. Imposing supersymmetry further constrains the Lagrangian terms. Enough supercharges can completely fix the theory. The field content we consider in this thesis is comprised of spin $\leq 1$ fields only, meaning scalars, spinors and vectors. A gauge theory is an additional way to define a QFT where the scalar and spinor fields are coupled to a massless vector via the covariant derivative. This requires the vector to be in the adjoint representation of the gauge group, and a singlet of the global symmetry. Note that only gauge equivalent states are physical and the gauge group describes a redundancy and not a symmetry. Gauge symmetry is the common but misleading name of this mechanism. Considering massless vectors not as part of a gauge theory, we find that the physical states are only Lorentz invariant when looking at the gauge equivalent configurations of the field [37]. Such a vector can be a background (non-dynamical) field, but requiring it to be dynamical and interacting while retaining Lorentz symmetry, causality and unitarity reverts the theory back to a gauge theory (and the vector becomes a proper gauge field) [38][1]. Therefore, interacting vectors are gauge fields. In supersymmetry, the building blocks we described are generalized to supermultiplets that will be defined for 3d theories in the following sections. Note we do not care for massive fields/multiplets because they do not participate in the theories of the fixed points.

As explained in the previous section, we can calculate the DCM using the supersymmetric index. The index is independent of the energy scale (RG flow) and can be calculated in a weakly coupled UV fixed point where the theory is free (in 3d all gauge theories are UV free, unlike the 4d case where it depends on the gauge group and matter content) [14]. In the free fixed point there are no interactions, so it is intuitive that the index requires less information compared to the full QFT. In essence, all that is required to know to calculate the index is the field content and symmetries (global and gauge). The form of the Lagrangian is irrelevant to the calculation. In the following sections we shall collect the building blocks required to calculate the index of 3d theories, later to be used to build the index of the SSQ theories. The definition and derivation of the 4d and 3d indices as a trace formula appears in Appendix B. Note that unlike in the 4d case, the scope of the trace formula is limited for the 3d index and the full calculation can only be done using the localization method, as explained in Appendix B.3.

### 3.2.1 $\mathcal{N} = 2$ **multiplets**

In $\mathcal{N} = 2$ theories the building blocks are the $\mathcal{N} = 2$ chiral multiplet and the $\mathcal{N} = 2$ vector multiplet. The index of an $\mathcal{N} = 2$ chiral multiplet with R-charge $r$ [39][16] is[1]

$$I_\chi(r, z, m; q) = \left(\frac{q^{\frac{1-r}{2}}}{z}\right)^{\frac{|m|}{2}} \frac{\left(q^{1-\frac{r}{2}+\frac{|m|}{2}}(-1)^{-m} z^{-1}; q\right)}{\left(q^{\frac{r}{2}+\frac{|m|}{2}}(-1)^m z; q\right)} \quad (3.3)$$

where

$$(z; q) \equiv \prod_{l=0}^{\infty} \left(1 - zq^l\right) \quad (3.4)$$

is the $q$-Pochhammer symbol. Here, $q$ is the fugacity related to the Lorentz symmetry and $U(1)$ R-symmetry, $z$ is a placeholder for global symmetry fugacities, and $m$ a placeholder for the corresponding magnetic monopole charges (also called GNO charges [40]). For a global symmetry group $G$ there are $\text{rank}(G)$ maximal torus variables $\mathbf{z}$, and additional $\text{rank}(G)$ monopole charge variables $\mathbf{m}$. The index for a general representation $R$ is given by

$$I_\chi(r, \mathbf{z}, \mathbf{m}; q) = \prod_{\mu \in \text{wt}(R)} I_\chi(r, \mathbf{z}^\mu, \mu(m); q) \quad (3.5)$$

where $\mu$ are the weights of the representation. See Appendices A.1 and B for the definitions of $\mathbf{z}^\mu$ and $\mu(m)$ and further details. It is worth mentioning here that fugacities for discrete global symmetries can also be defined [41], but we not shall deal with them in this thesis. The calculation of the $\mathcal{N} = 2$ chiral multiplet index in the zero-monopole sector $m = 0$ can be performed using the trace formula, see Appendix B.3.

Unfortunately, the index of the $\mathcal{N} = 2$ vector multiplet cannot be calculated using the trace formula, as we also explain in Appendix B.3. The vector contribution can only be calculated using localization, and the result is that for a general gauge group $G$ it can be written in the form [39][16]

$$I_V(\mathbf{z}, \mathbf{m}; q) = \prod_{\alpha \in \text{rt}(G)} q^{-\frac{1}{4}|\alpha(m)|} \left(1 - (-1)^{\alpha(m)} \mathbf{z}^\alpha q^{\frac{1}{2}|\alpha(m)|}\right) \quad (3.6)$$

where $\alpha$ are the roots of the gauge group. The definitions of $\mathbf{z}^\alpha$ and $\alpha(m)$ are the same as used above for the weights, as defined in Appendices A.1 and B.3.1. Note there is no contribution for the abelian $U(1)$ group (no roots). A summary of roots for the classical Lie groups can be found in [42] (p. 362). For example, the roots of $SU(N)$ are given by $\vec{\alpha} = \hat{e}_i - \hat{e}_j$ for $i \neq j = 1, ..., N$, where $\hat{e}_i$ are unit vectors in $N$-dimensional

---
[1] Eq. (5.20) in [16] has a mistake in the prefactor, apart from that both expressions are equivalent after the redefinition $z \to (-1)^m z$.

space. This gives $\alpha(m) = m_i - m_j$. Plugging this into Eq. 3.6 gives:

$$I_V(\mathbf{z}, \mathbf{m}; q) = \prod_{i \neq j=1}^{N} q^{-\frac{|m_i - m_j|}{4}} \left(1 - (-1)^{|m_i - m_j|} \frac{z_i}{z_j} q^{\frac{|m_i - m_j|}{2}}\right) \qquad (3.7)$$

Note that specifically for $SU(N)$ the rank is $N-1$ but it is convenient to use $N$ variables (and an $N$-dimensional space for the weights and roots) with the additional constraints $\prod_{j=1}^{N} z_j = 1$ and $\sum_{j=1}^{N} m_j = 0$ (see Appendix B.3.1). For $U(N) \sim U(1) \times SU(N)$ of rank $N$ it is the same result just without the constraints for $\mathbf{z}$ and $\mathbf{m}$. From here on we do not explicitly refer to the constraints, it is to be understood depending on the group.

The vector contribution can also be calculated as if it is an adjoint chiral multiplet with $r = 2$ [16], with the zero weights (Cartan generators) omitted from the adjoint weight system (same as the root system, see Appendix B.3.1). Remember that we consider vectors only as gauge fields and therefore they must appear in the adjoint representation of the group.

For example, the adjoint character of $SU(N)$ is:

$$\chi_{\text{adj}}^{SU(N)}(\mathbf{z}) = \sum_{1 \leq i \neq j \leq N} z_i z_j^{-1} + (N-1) \qquad (3.8)$$

For $U(N)$ the adjoint character is $\chi_{\text{adj}}^{U(N)}(\mathbf{z}) = \chi_{\text{adj}}^{SU(N)}(\mathbf{z}) + 1$, but we are going to descard the zero weights so the result is the same for $SU(N)$ and $U(N)$. Discard and plug into the adjoint chiral index with $r = 2$:

$$\begin{aligned}
I_V^{SU(N)}(\mathbf{z}, \mathbf{m}; q) &= I_V^{U(N)}(\mathbf{z}, \mathbf{m}; q) \\
&= \prod_{i \neq j=1}^{N} I_\chi\left(r = 2, \frac{z_i}{z_j}, m_i - m_j; q\right) \\
&= \prod_{i \neq j=1}^{N} \left(\frac{q^{\frac{1}{2}}}{z_i/z_j}\right)^{-\frac{|m_i - m_j|}{2}} \frac{\prod_{l=0}^{\infty}\left(1 - q^{\frac{|m_i - m_j|}{2}} (-1)^{-(m_i - m_j)} \left(\frac{z_i}{z_j}\right)^{-1} q^l\right)}{\prod_{l=0}^{\infty}\left(1 - q^{1 + \frac{|m_i - m_j|}{2}} (-1)^{m_i - m_j} \frac{z_i}{z_j} q^l\right)} \\
&= \prod_{i \neq j=1}^{N} q^{-\frac{|m_i - m_j|}{4}} \left(1 - (-1)^{m_i - m_j} \frac{z_i}{z_j} q^{\frac{|m_i - m_j|}{2}}\right)
\end{aligned} \qquad (3.9)$$

Notice that due to the character structure the terms in the numerator and denominator cancel out almost completely, and only the numerator for $l = 0$ remains. We can see the result matches Eq. 3.7 as advertised.

### 3.2.2 $\mathcal{N} = 4$ multiplets

Moving on to $\mathcal{N} = 4$ theories (like the SSQ theories) the building blocks are the $\mathcal{N} = 4$ hypermultiplet and the $\mathcal{N} = 4$ vector multiplet. In $\mathcal{N} = 4$ the R-symmetry is enhanced to $SO(4) \sim SU(2)_R \times SU(2)_C$ (rank 2), and therefore an additional fugacity will be required for the R-symmetry (see chapter 2). Since the R-symmetry generators $R_H, R_C$ appear in the index in the combinations $R_H \pm R_C$ only, the R-symmetry is described in terms of the $U(1) \times U(1)$ subgroup generators (see Appendix A.4). Therefore, an $\mathcal{N} = 4$ theory can be described as an $\mathcal{N} = 2$ theory (that has $U(1)$ R-symmetry) with an additional global symmetry $U(1)_t$ (whose fugacity is $t$) [16].

Also note that the enlarged supersymmetry in $\mathcal{N} = 4$ fixes the R-charges of the multiplets (will be verified with Z-extremization [32] in section 3.5). For a more elaborate derivation of the $\mathcal{N} = 4$ case see Appendix B.4.

The $\mathcal{N} = 4$ hypermultiplet is composed of two conjugate $\mathcal{N} = 2$ chiral multiplets. The index of the hypermultiplet therefore is the product of two conjugate $\mathcal{N} = 2$ chiral multiplets with $r = \frac{1}{2}$ and the additional fugacity $t$ is incorporated using the mapping $z \to zt^{\frac{1}{2}}$ (note that the conjugation does not apply to $t$) [11]:

$$
\begin{aligned}
I_{\mathrm{hyp}}(z, m; q, t) &= I_\chi\left(r = \frac{1}{2}, zt^{\frac{1}{2}}, m; q\right) I_\chi\left(r = \frac{1}{2}, z^{-1}t^{\frac{1}{2}}, -m; q\right) \\
&= \left(\frac{q^{\frac{1}{4}}}{zt^{\frac{1}{2}}}\right)^{\frac{|m|}{2}} \frac{\left(q^{\frac{3}{4}+\frac{|m|}{2}}(-1)^{-m} z^{-1} t^{-\frac{1}{2}}; q\right)}{\left(q^{\frac{1}{4}+\frac{|m|}{2}}(-1)^{m} zt^{\frac{1}{2}}; q\right)} \\
&\quad \cdot \left(\frac{q^{\frac{1}{4}}}{z^{-1}t^{\frac{1}{2}}}\right)^{\frac{|m|}{2}} \frac{\left(q^{\frac{3}{4}+\frac{|m|}{2}}(-1)^{m} zt^{-\frac{1}{2}}; q\right)}{\left(q^{\frac{1}{4}+\frac{|m|}{2}}(-1)^{-m} z^{-1}t^{\frac{1}{2}}; q\right)} \\
&= \left(\frac{q^{\frac{1}{2}}}{t}\right)^{\frac{|m|}{2}} \frac{\left(t^{-\frac{1}{2}} q^{\frac{3}{4}+\frac{|m|}{2}} z^{\pm 1}; q\right)}{\left(t^{\frac{1}{2}} q^{\frac{1}{4}+\frac{|m|}{2}} z^{\pm 1}; q\right)}
\end{aligned}
\quad (3.10)
$$

Above we used the notation $f(z^{\pm 1}) \equiv f(z) f(z^{-1})$.

The $\mathcal{N} = 4$ vector multiplet is composed of an $\mathcal{N} = 2$ chiral multiplet and an $\mathcal{N} = 2$ vector multiplet, both in the adjoint representation of the gauge group. The $\mathcal{N} = 2$ chiral multiplet (we denote by the subscript $\chi v$) contributes with $r = 1$ and with the mapping $z \to zt^{-1}$:

$$
\begin{aligned}
I_{\chi v}(z, m; q, t) &= I_\chi(r = 1, zt, m; q) \\
&= \left(\frac{1}{zt^{-1}}\right)^{\frac{|m|}{2}} \frac{\left(q^{\frac{1}{2}+\frac{|m|}{2}}(-1)^{-m} z^{-1} t; q\right)}{\left(q^{\frac{1}{2}+\frac{|m|}{2}}(-1)^{m} zt^{-1}; q\right)}
\end{aligned}
\quad (3.11)
$$

For $SU(N)$ we plug in the character of the adjoint (again, $t$ is unaffected):

$$I_{\chi v}^{SU(N)}(\mathbf{z}, \mathbf{m}; q, t) = I_{\chi v}(1, 0; q, t)^{N-1} \prod_{i \neq j=1}^{N} I_{\chi v}\left(\frac{z_i}{z_j}, m_i - m_j; q, t\right)$$

$$= \left(\frac{\left(q^{\frac{1}{2}}t; q\right)}{\left(q^{\frac{1}{2}}t^{-1}; q\right)}\right)^{N-1}$$

$$\cdot \prod_{i \neq j=1}^{N} \left(\frac{1}{t}\right)^{-\frac{|m_i - m_j|}{2}} \frac{\left(q^{\frac{1}{2} + \frac{|m_i - m_j|}{2}}(-1)^{-(m_i - m_j)}\left(\frac{z_i}{z_j}\right)^{-1} t; q\right)}{\left(q^{\frac{1}{2} + \frac{|m_i - m_j|}{2}}(-1)^{m_i - m_j} \frac{z_i}{z_j} t^{-1}; q\right)}$$

(3.12)

For $U(N)$ there would be an additional factor of $I_{\chi v}(1, 0; q, t)$.

For a more rigorous calculation of the index of the $\mathcal{N} = 2$ chiral multiplets that are a part of the $\mathcal{N} = 4$ multiplets in the zero-monopole sector $m = 0$ using the trace formula see Appendix B.4.

The $\mathcal{N} = 2$ vector multiplet contribution is the same as in the previous section, because the vector is charged only with the gauge group and not with global symmetries (so it does not depend on $t$). The result for $SU(N)$ or $U(N)$ was Eq. 3.7:

$$I_V^{SU(N)}(\mathbf{z}, \mathbf{m}; q) = \prod_{i \neq j=1}^{N} \left(q^{\frac{1}{2}}\right)^{-\frac{|m_i - m_j|}{2}} \left(1 - (-1)^{m_i - m_j} \frac{z_i}{z_j} q^{\frac{|m_i - m_j|}{2}}\right) \quad (3.13)$$

The combined index of the $\mathcal{N} = 4$ vector multiplet for $SU(N)$ is therefore:

$$I_{V, \mathcal{N}=4}^{SU(N)}(\mathbf{z}, \mathbf{m}; q, t) = I_{\chi v}^{SU(N)}(\mathbf{z}, \mathbf{m}; q, t) I_V^{SU(N)}(\mathbf{z}, \mathbf{m}; q)$$

$$= \left(\frac{\left(q^{\frac{1}{2}}t; q\right)}{\left(q^{\frac{1}{2}}t^{-1}; q\right)}\right)^{N-1} \prod_{i \neq j=1}^{N} \left(\frac{q^{\frac{1}{2}}}{t}\right)^{-\frac{|m_i - m_j|}{2}} \left(1 - (-1)^{m_i - m_j} \frac{z_i}{z_j} q^{\frac{|m_i - m_j|}{2}}\right)$$

$$\cdot \frac{\left(q^{\frac{1}{2} + \frac{|m_i - m_j|}{2}}(-1)^{-(m_i - m_j)}\left(\frac{z_i}{z_j}\right)^{-1} t; q\right)}{\left(q^{\frac{1}{2} + \frac{|m_i - m_j|}{2}}(-1)^{m_i - m_j} \frac{z_i}{z_j} t^{-1}; q\right)}$$

(3.14)

For $U(N)$ the result is $I_{V, \mathcal{N}=4}^{U(N)} = I_{\chi v}^{U(N)} I_V^{U(N)}$ which amounts to an additional factor of $I_{\chi v}(1, 0; q, t)$.

Note that for $\mathcal{N} = 4$ theories the $(-1)^m$ factors can be omitted from the index expressions for simplicity (see footnote 6 in [11]). The way to see this is to redefine $z \to (-1)^m z$. In the $\mathcal{N}=4$ vector multiplet index $I_{V,\mathcal{N}=4}$ (Eq. 3.14) this redefinition trivially hides the $(-1)^m$ factors. In the $\mathcal{N}=2$ chiral multiplet index $I_\chi$ (Eq. 3.3) this redefinition hides most of the $(-1)^m$ factors, but leaves a prefactor $e^{-i\pi m^2/2}$ (a form similar to the one in [16]). The hypermultiplet index $I_{\text{hyp}}$ (Eq. 3.10) is composed of two conjugate $\mathcal{N}=2$ chiral multiplets and therefore these factors cancel out.

Also note that each $\mathcal{N}=4$ multiplet has a "twisted" counterpart where the $SU(2)_R, SU(2)_C$ factors of the R-symmetry are swapped. For the hypermultiplet, this means mapping $t \to t^{-1}$ to get the index of the twisted hypermultiplet [11].

### 3.2.3 Gauge theory

Additionally to the building blocks defined in the previous section, here we describe how to treat the gauge part of the theory. The fugacities related to a gauge group cannot remain in the final expression of the index since they correspond to redundant degrees of freedom and do not describe a global symmetry group. As explained in Appendix B.2, the supersymmetric index of a gauge theory receives contributions only from gauge-invariant states, which requires to integrate the gauge group maximal torus variables over the group using the Haar measure of the group $\mathrm{d}\mu_G$ (see Appendices A.2 and B.2). Therefore, given a theory with global symmetry group $G$ and a corresponding supersymmetric index $I(\mathbf{z})$, the index of the gauge theory (after gauging the global symmetry group) is:

$$I_{\text{gauge}} = \int_G \mathrm{d}\mu_G(\mathbf{z}) I(\mathbf{z}) \tag{3.15}$$

This is indeed how the index of a gauge theory is calculated in the 4d case [15]. $I(\mathbf{z})$ already contains the contributions of all the fields, including the vector (gauge field). In odd dimensions (3d and 5d) the situation is different, because there are non-trivial gauge field configurations that contribute and can alter the above prescription. By non-trivial we mean their states cannot be created using operations of polynomials in the gauge field $\mathbf{A}$ or field strength tensor $\mathbf{F}$ operators.

The Levi-Civita antisymmetric tensor has $d$ indices in $d$ spacetime dimensions, and it allows to write a conserved current out of gauge fields in odd dimensions only. In 3d every abelian factor of the gauge group allows to write a gauge-invariant conserved current $\mathbf{j} = \text{tr}(\star \mathbf{F})$ (in differential form) [11][16][43] and in 5d every non-abelian factor of the gauge group allows to write a gauge-invariant conserved current $\mathbf{j} = \text{tr}(\star \mathbf{F} \wedge \mathbf{F})$ [44]. The currents are spacetime vectors and the conservation means $\mathrm{d}\mathbf{j} = 0$. These conserved currents are associated with a topological global symmetry $U(1)_J$ (not evident in the Lagrangian formulation) whose charged objects are the non-trivial gauge field configurations, being monopoles in 3d [11][16][43][41] and instantons in 5d [44][35]. It turns out that these non-trivial configurations contribute to the supersymmetric index in a non-trivial way, calculated using localization.

Here we write down the result of the computation for 3d [16]. In the non-zero monopole sectors the gauge group is (partially) broken. All (inequivalent) monopole configurations need to be summed, for which we use the notation

$$\sum_{\mathbf{m}} \equiv \left( \prod_{i=1}^{\text{rank}(G)} \sum_{m_i} \right) \frac{1}{|W(\mathbf{m})|} \quad (3.16)$$

where $|W(\mathbf{m})|$ is the dimension of the residual Weyl group in each monopole configuration (due to the broken gauge group) [11][39]. If we sum all the combinations $\{m_i\} \in \mathbb{Z}$ for all $i$, we are overcounting gauge-equivalent combinations. However, this calculation is still valid, given we use the original unbroken Weyl group dimension $|W|$ (see Appendix A.2) for all the terms in the sum.

Yet another difference in the 3d computation involves the integration over the gauge group. While in the 4d case the integration is weighed by the Haar measure $\int_G d\mu_G(\mathbf{z})$, in 3d it is not as it will be compensated by a different contribution, so we write the integration as $\oint \frac{d\mathbf{z}}{2\pi i \cdot \mathbf{z}}$ (for the shorthand notation see Appendix A.2). The index contribution we wrote for the vector $I_V$ (Eq. 3.6) is composed both from the contribution of the Haar measure (in different monopole sectors) and of the gaugino (the spinor of the $\mathcal{N}=2$ vector multiplet). We can see that the zero monopole sector ($m_i - m_j = 0$ for all $i,j$) exactly reproduces the non-abelian contribution to the Haar measure. For non-zero monopole sectors ($m_i - m_j \neq 0$ for some $i,j$) the contribution also depends on $q$, which indicates that it comes from the gauginos (unlike the Haar measure which depends only on the fugacities of the gauge group $\mathbf{z}$).

In total, the index of a 3d gauge theory is given by

$$I_{\text{gauge}} = \sum_{\mathbf{m}} \oint \frac{d\mathbf{z}}{2\pi i \cdot \mathbf{z}} I_V(\mathbf{z}, \mathbf{m}) I_{\text{mat}}(\mathbf{z}, \mathbf{m}) \quad (3.17)$$

where $I_{\text{mat}}(\mathbf{z}, \mathbf{m})$ (mat stands for matter) is the index of theory prior to gauging, meaning a theory containing some chiral multiplets (without vectors).

Thus far we described how the gauge degrees of freedom are taken care of in 3d. However, gauge theories can still have global symmetries. If the matter contribution has an additional (ungauged) global symmetry described with fugacities $\mathbf{a}$, that global symmetry remains after gauging:

$$I_{\text{gauge}}(\mathbf{a}) = \sum_{\mathbf{m}} \oint \frac{d\mathbf{z}}{2\pi i \cdot \mathbf{z}} I_V(\mathbf{z}) I_{\text{mat}}(\mathbf{z}, \mathbf{a}) \quad (3.18)$$

For the topological global symmetry $U(1)_J$ (if exists) we define a (single) fugacity $b$. This fugacity is charged with the fluxes of the gauge group on $\mathbb{S}^2$ (magnetic monopole charges) $\mathbf{m}$ and contributes to the index in the form $b^{\mathbf{m}} = \prod_{i=1}^{\text{rank}(G)} b^{m_i} = b^{\sum_{i=1}^{\text{rank}(G)} m_i}$.

Also, a Fayet-Iliopoulos (FI) term can be added to the Lagrangian, which is gauge invariant only for the abelian factor of the gauge group [2]. It can also be thought of as a BF coupling term of a background twisted $\mathcal{N}=4$ vector multiplet to the global $U(1)_J$ symmetry [45][11]. The flux $n$ of this background field couples to the $U(1)$ factor of the gauge group (the $U(1)$ combination of the gauge group fugacities). Therefore, for $U(N)$ the coupling is of the form $\mathbf{z}^n = \prod_{i=1}^{\text{rank}(G)} z_i^n$ [11] and for pure non-abelian groups there is no coupling. We will only look at the zero flux case in this thesis, meaning $n = 0$.

Lastly, the same reason that the topological conserved current exists in odd dimensions also allows to write a Chern-Simons (CS) term in the Lagrangian [46]. A CS term of level $k$ couples to the gauge fugacities $\mathbf{z}$ as $\mathbf{z}^{k\mathbf{m}} = \prod_{i=1}^{\text{rank}(G)} z_i^{km_i}$ [47]. The maximal supersymmetry the CS term is consistent with is $\mathcal{N} = 3$ [48][16] (unless there is no Yang-Mills term [49][11]). The SSQ theories we consider in this thesis are $\mathcal{N} = 4$ and therefore not consistent with CS terms, meaning $k = 0$.

Note that it is possible to generalize the treatment we did above and define an additional flux for every global/flavor symmetry fugacity (and even an additional CS contribution corresponding to that flux). This allows to extract more information out of the index, and provide a stronger test of dualities [45][47]. However, in this thesis we will not allow these additional fluxes because our ability to extract the marginal operators (to eventually calculate the DCM) relies on the standard supersymmetric index [17][18]. What can be extracted from the generalized supersymmetric index is an interesting open question.

In total, the index of a general 3d gauge theory is given by:

$$I_{\text{gauge}}(b, n, k, \mathbf{a}) = \sum_{\mathbf{m}} \oint \frac{d\mathbf{z}}{2\pi i \cdot \mathbf{z}} b^{\mathbf{m}} \mathbf{z}^{k\mathbf{m}+n} I_V(\mathbf{z}, \mathbf{m}) I_{\text{mat}}(\mathbf{z}, \mathbf{m}, \mathbf{a}) \qquad (3.19)$$

### 3.2.4 Numerical methods

The 3d indices involve $q$-Pochhammer symbols which are infinite products. Since we will ultimately be interested in the expansion of the index in $q$ to a finite order, and the $q$-Pochhammer symbol (Eq. 3.4) involves a product of increasing powers of $q$, we can limit the product to a finite number of terms $(z; q) \equiv \prod_{l=0}^{q_{max}} \left(1 - zq^l\right)$ where $q_{max}$ is some positive integer. In this thesis we found that $q_{max} = 1$ was a converged value for index expansion to first order in $q$.

For gauge theories we need to consider how to numerically treat two non-trivial elements: the sum of different gauge group fluxes and the contour integration.

For the sum on gauge group fluxes, we need to cover all flux combinations that contribute to the index. Therefore, we consider all $\mathbf{m} = \left(m_1, ..., m_{\text{rank}(G)}\right)$ such that each $|m_i| \leq m_{max}$ for some integer $m_{max}$. In this thesis we found $m_{max} = 2$ was a converged value for all the gauge theories we analyzed (in the expansion to first order in $q$). Another helpful feature is using the symmetry of the Weyl group, of which the simplest equivalence we can use for all gauge groups is the symmetry under exchange of fluxes (exchanging $m_i$ and $m_j$ in $\mathbf{m}$ for any $i, j$). This means we only need to calculate the index contributions of flux combinations that are of independent equivalence classes, and multiply by the multiplicity of the combination (since we normalize by the dimension of the unbroken Weyl group, see Appendix A.2). For example, for a rank 2 gauge group and $m_{max} = 1$ the equivalence classes will be $\mathbf{m} = (0,0), (1,0), (-1,0), (1,1), (1,-1)$, and their multiplicities are $1, 2, 2, 2, 2$, respectively. For large rank gauge groups this will allow an exponential decrease in the number of flux combinations that need to be calculated.

As for the contour integrals, solving them is generally hard. However, since we are interested in expansions in $q$, we can simplify by first expanding the integrand in $q$ and only then performing the contour integration. By expanding the integrand in $q$ we transform the index from a product form (in the numerator and denominator) to a series form where each term is given by a product of different fugacities $\sum q^{\#} t^{\#} \mathbf{a}^{\#} \mathbf{z}^{\#}$. This simplified form can be solved using the Residue theorem by taking only the terms proportional to $\mathbf{z}^{-1} = \prod_{i=1}^{\text{rank}(G)} z_i^{-1}$ (the fugacities associated with the gauge group). To achieve a converged value for the expansion of the integral itself to first order in $q$ for a gauge theory with a single gauge group, it is trivial that we need to expand the integrand to first order in $q$. For a more complicated case of a gauge theory composed of several gauge groups (as will be the case for the $T[SU(N)]$ theories in section 3.3), the expansion of the internal gauge groups to first order in $q$ might not be enough to reach convergence.

## 3.3 The $T[SU(N)]$ linear quiver theories

Before we define the SSQ theories, we need to define another building block which are the $T[SU(N)]$ linear quiver theories (which are $\mathcal{N} = 4$ supersymmetric). Even before that, we define a bifundamental hypermultiplet as a hypermultiplet that is in the fundamental representation of one group, and also in the anti-fundamental representation of a second group. The $T[SU(N)]$ theories were previously described in [50], a systematic construction of the theory for integer $N \geq 2$:

- Start with bifundamental hypermultiplet of $SU(N)$ and $U(N-1)$ groups. Gauge the $U(N-1)$ group. In gauging an adjoint $\mathcal{N} = 4$ vector multiplet of $U(N-1)$ appears.

- Add another bifundamental hypermultiplet of (the previous) $U(N-1)$ group and of a new $U(N-2)$ group. Again, gauge the $U(N-2)$ group (which brings an accompanying $\mathcal{N} = 4$ vector multiplet of $U(N-2)$).

- Repeat for smaller and smaller gauge groups until a bifundamental hypermultiplet of $U(2)$ and $U(1)$ groups is added and the $U(1)$ is gauged.

In total this is a gauge theory with gauge group $U(N-1) \times U(N-2) \times ... \times U(1)$ and $N-1$ different bifundamental hypermultiplets.

A diagramatic description:

Figure 3.1: Quiver diagram of the $T[SU(N)]$ theory

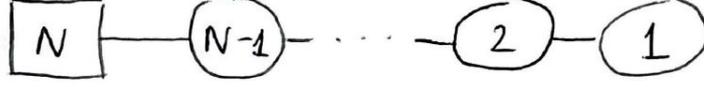

In the diagram, a square labeled $N$ represents an $SU(N)$ global symmetry group, a circle labeled $N$ representation a $U(N)$ gauge group and the lines connecting two groups represents a bifundamental hypermultiplet of the groups.

The character of the fundamental representation of $U(N)$ is $\chi_{\text{fund}}^{U(N)} = \sum_{i=1}^{N} z_i$ (and of $SU(N)$ is the same with the constraint $\prod_{i=1}^{N} z_i = 1$). For the anti-fundamental representation transform every variable $z \to z^{-1}$. Note that for $SU(2)$ both representations are equivalent since the group is real. Define the set of maximal torus variables of each group $\left\{\mathbf{z}^{(k)}\right\}_{k=1}^{N}$, where $k = 1$ refers to the group $U(1)$ and $k = N$ refers to the leftmost group $SU(N)$. Therefore, the index of a bifundamental hypermultiplet associated with the gauge groups $U(k), U(k+1)$ (for $1 \leq k \leq N-2$):

$$I_{\text{bfh}}^{(k)}\left(\mathbf{z}^{(k)}, \mathbf{m}^{(k)}, \mathbf{z}^{(k+1)}, \mathbf{m}^{(k+1)}; q, t\right) = \prod_{j=1}^{k+1} \prod_{i=1}^{k} I_{\text{hyp}}\left(z_i^{(k)}/z_j^{(k+1)}, m_i^{(k)} - m_j^{(k+1)}; q, t\right) \tag{3.20}$$

For the case $k = N-1$, the bifundamental hypermultiplet is associated with the gauge group $U(N-1)$ and the un-gauged group $SU(N)$ and so we apply the constraints $\prod_{j=1}^{N} z_j^{(N)} = 1$ and $\sum_{j=1}^{N} m_j^{(N)} = 0$ (see previous secction). In the case $N = 2$ there is only one bifundamental hypermultiplet and gauge group $U(1)$ (this case is discussed in [11]).

The $\mathcal{N} = 4$ vector contribution for each gauge group $U(k)$ is $I_{V,\mathcal{N}=4}^{U(N)}\left(\mathbf{z}^{(k)}, \mathbf{m}^{(k)}; q, t\right)$.

As explained in the previous section, in each integration of a gauge group $U(k)$ it is possible to add a flux $n$ corresponding the FI term in the $U(1)$ factors of the gauge groups, and a fugacity $b$ for the monopole charge of the topological $U(1)$ group. For the theory $T[SU(N)]$ this gives new parameters $\mathbf{n} \equiv \left\{n^{(k)}\right\}_{k=1}^{N-1}$ and $\mathbf{b} \equiv \left\{b^{(k)}\right\}_{k=1}^{N-1}$. We take the fluxes $\mathbf{n}$ to be zero in this thesis. Also, the Weyl group dimension $|W|$ for the gauge group $U(k)$ is $k!$.

Finally, the index of $T[SU(N)]$ can be computed by integrating all the gauge groups recursively from $k = 0$ to $k = N-1$ (using the gauge theory index Eq. 3.19 from the

previous section):

$$I_{T[SU(N)]}\left(\mathbf{z}^{(N)}, \mathbf{m}^{(N)}, \mathbf{b}; q, t\right) \equiv I_{N-1}^{\text{rec}}$$
$$I_k^{\text{rec}} = \sum_{\mathbf{m}^{(k)}} \oint \frac{d\mathbf{z}^{(k)}}{2\pi i \cdot \mathbf{z}^{(k)}} \left(b^{(k)}\right)^{\mathbf{m}^{(k)}} I_{V,\mathcal{N}=4}^{U(N)}\left(\mathbf{z}^{(k)}, \mathbf{m}^{(k)}; q, t\right)$$
$$\cdot I_{\text{bfh}}^{(k)}\left(\mathbf{z}^{(k)}, \mathbf{m}^{(k)}, \mathbf{z}^{(k+1)}, \mathbf{m}^{(k+1)}; q, t\right) \cdot I_{k-1}^{\text{rec}}$$
$$I_0 = 1 \tag{3.21}$$

The global topological symmetry $U(1)_J^{N-1}$ enhances in the IR to $SU(N)_J$ [50][11][41]. This is a conjecture that can be checked, since it means the set of fugacities **b** are expected to appear as characters of $SU(N)$ in the index.

### 3.3.1 Calculation of the index

We calculated the index of $T[SU(N)]$ for $2 \leq N \leq 5$ for zero fluxes $\mathbf{m}^{(N)}$ (and for $2 \leq N \leq 6$ for non-zero fluxes $\mathbf{m}^{(N)}$) to first order in $q$. The indices for $N \geq 6$ for zero fluxes proved numerically prohibitive to compute. Using the calculated index, we were able to find the transformation needed to make the **b** fugacities appear as characters of $SU(N)$, as conjectured. See Appendix A.3.1 for more details.

We use the notation $\chi$ for the characters of $SU(N)_J$ (**b** fugacities) and $\tilde{\chi}$ for the characters of the second $SU(N)$ ($\mathbf{z}^{(N)}$ fugacities). The index of the zero flux sector $\mathbf{m}^{(N)} = 0$ up to first order in $q$ is

$$I_{T[SU(N)]} = 1 + q^{\frac{1}{2}}\left(\chi_{\text{adj}}^{SU(N)}\frac{1}{t} + \tilde{\chi}_{\text{adj}}^{SU(N)}t\right)$$
$$+ q\left(A(N)\frac{1}{t^2} + \tilde{A}(N)t^2 + \chi_{\text{adj}}^{SU(N)}\tilde{\chi}_{\text{adj}}^{SU(N)}\delta_{N\geq 3}\underbrace{-1 - \tilde{\chi}_{\text{adj}}^{SU(N)} - \chi_{\text{adj}}^{SU(N)}}_{\text{conserved currents}}\right) \tag{3.22}$$

where the adjoint character of $SU(N)$ in terms of Dynkin labels [51] is given by

$$\chi_{\text{adj}}^{SU(N)} = \begin{cases} \chi_{[2]}^{SU(2)} & N = 2 \\ \chi_{[1,0,\ldots,0,1]}^{SU(N)} & N \geq 3 \end{cases} = \begin{cases} \chi_{\mathbf{3}}^{SU(2)} & N = 2 \\ \chi_{\mathbf{8}}^{SU(3)} & N = 3 \\ \chi_{\mathbf{15}}^{SU(4)} & N = 4 \\ \chi_{\mathbf{24}}^{SU(5)} & N = 5 \end{cases} \tag{3.23}$$

and $A(N)$ is a shorthand notation for several characters:

$$A(N) \equiv \begin{cases} \chi^{SU(2)}_{[4]} & N = 2 \\ \chi^{SU(3)}_{[1,1]} + \chi^{SU(3)}_{[2,2]} & N = 3 \\ \chi^{SU(4)}_{[1,0,1]} + \chi^{SU(4)}_{[0,2,0]} + \chi^{SU(4)}_{[2,0,2]} & N = 4 \\ \chi^{SU(5)}_{[1,0,0,1]} + \chi^{SU(5)}_{[0,1,1,0]} + \chi^{SU(5)}_{[2,0,0,2]} & N = 5 \end{cases}$$
$$= \begin{cases} \chi^{SU(2)}_{\mathbf{5}} & N = 2 \\ \chi^{SU(3)}_{\mathbf{8}} + \chi^{SU(3)}_{\mathbf{27}} & N = 3 \\ \chi^{SU(4)}_{\mathbf{15}} + \chi^{SU(4)}_{\mathbf{20'}} + \chi^{SU(4)}_{\mathbf{84}} & N = 4 \\ \chi^{SU(5)}_{\mathbf{24}} + \chi^{SU(5)}_{\mathbf{75}} + \chi^{SU(5)}_{\mathbf{200}} & N = 5 \end{cases} \quad (3.24)$$

$\tilde{A}(N)$ is the same using the $\tilde{\chi}$ characters. By examining the form of the Dynkin labels of the representations contributing to $A(N)$, we might try to guess the pattern and conjecture the form for $N \geq 6$. In this case the pattern is not obvious but the representations given by the Dynkin labels $\chi^{SU(N)}_{[1,0,...,0,1]} + \chi^{SU(N)}_{[2,0,...,0,2]}$ will probably remain for high $N$.

The indices of non-zero flux sectors $\mathbf{m}^{(N)} \neq 0$ are given by

$$I_{T[SU(N)]} = \begin{cases} \frac{q^{\frac{1}{2}}}{t}\chi^{SU(2)}_{\mathbf{3}} + \frac{q}{t^2}\left(\chi^{SU(2)}_{\mathbf{5}} - 1\right) & N = 2, \mathbf{m}^{(N)} = (1) \\ \frac{q}{t^2}\chi^{SU(2)}_{\mathbf{5}} & N = 2, \mathbf{m}^{(N)} = (2) \\ \frac{q}{t^2}\chi^{SU(3)}_{\mathbf{8}} & N = 3, \mathbf{m}^{(N)} = (1,0), (1,-1) \end{cases} \quad (3.25)$$

where in each case we indicated $N$ and the flux equivalence classes $\mathbf{m}^{(N)}$ (see chapter 3.2.4). All other non-zero flux combinations for $N = 2, 3$ and any non-zero flux configurations for $N = 4, 5, 6$ are zero. We conjecture it remains zero for $N > 6$ as well (to first order in $q$).

Note the characters of the flavor $SU(N)$ symmetry (given in terms of the $\mathbf{z}^{(N)}$ fugacities) do not appear since the symmetry was broken by the non-zero fluxes. Also note that in the $N = 5$ case, achieving convergence of the expansion to first order in $q$ required the indices of the internal gauge groups ($U(N)$ for $N \leq 4$) to be expanded to second order in $q$.

The calculations and subsequent fitting of the fugacities with characters of the global symmetry groups were performed in an accompanying Mathematica notebook [52][53].

## 3.4 The star-shaped-quiver theories

Finally we are ready to define the SSQ theories (which are $\mathcal{N} = 4$ supersymmetric). These theories were previously described in [13], and are constructed as follows:

- Take $s$ "legs" made of $T[SU(N)]$ theories, identify the $SU(N)$ global symmetry of all the legs and gauge it as $SU(N)/\mathbb{Z}_N$. This central node with the legs sticking out is the origin of the star-shaped-quiver name. Note that an adjoint $\mathcal{N}=4$ vector multiplet of $SU(N)$ appears in the process of gauging. We remind that the central node is described by the fugacities $\mathbf{z}^{(N)}$ and fluxes $\mathbf{m}^{(N)}$ from the previous section on $T[SU(N)]$ theories.

- Prior to gauging, add $g$ $\mathcal{N}=4$ hypermultiplets that are adjoints of the central node $SU(N)/\mathbb{Z}_N$ gauge group.

- The class of SSQ theories discussed in this thesis are defined using the parameters $s, g, N$ (which are integers such that $s, g \geq 0$ and $N \geq 2$). The $N=2$ case was previously discussed in [11].

A diagramatic description:

Figure 3.2: Quiver diagram of the SSQ theory with parameters $s, g, N$

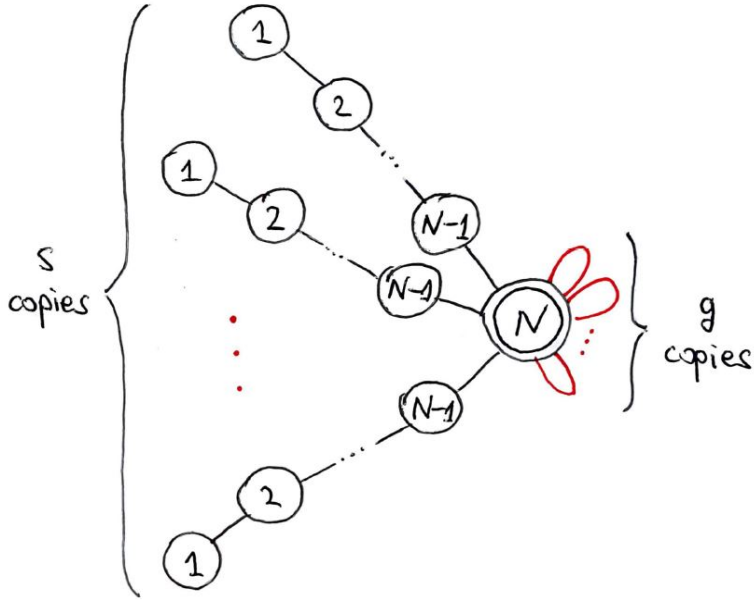

The definitions of what the circles and the lines connecting them represent are the same as in Fig 3.1, with the additional definitions that a double-lined circle represents an $SU(N)/\mathbb{Z}_N$ gauge group, and a line connecting a circle to itself (in the central node) represents an adjoint hypermultiplet of that group.

Notes:

- The index diverges for certain theories called "bad", which happens because the IR superconformal R-symmetry is not identified correctly [50][11]. This can happen for example when the R-symmetry is enhanced. Sometimes a "bad" theory can have a dual "good" theory [11]. Such is the case of the $s=0, g=1, N=2$ theory which is known to be "bad", however it has a dual "good" theory, the $\mathcal{N} = 8$ ABJM theory [49].

- As discussed in the introduction to the thesis, we are interested in the mirror duals of the class $\mathfrak{s}$ theories, which are the SSQ theories with the gauge group on the central node conjectured to be $SU(N)/\mathbb{Z}_N$ rather than $SU(N)$ [11][41]. The difference between the two cases is in the magnetic monopole flux configurations that need to be taken into account. Since we are interested in the marginal operators only (for the DCM), the difference between the two different gauge theories is expected to disappear for high enough $g$. This is because the marginal operators appear in power $q$ in the index expansion, and higher $g$ pushes the monopole contributions to higher order in $q$. We deal only with the $SU(N)$ case in this thesis.

Regarding the global symmetry:

- The $g$ $\mathcal{N} = 4$ hypermultiplets are indistinguishable and each is composed of two $\mathcal{N} = 2$ chiral multiplets in conjugate representations of $SU(N)$, so we have two sets of $g$ $\mathcal{N} = 2$ chiral multiplets. There is a $SU(g)$ rotation symmetry for the chiral multiplets in one of the sets, and a $U(1)$ rotation symmetry between the two sets. The second set does not have a $SU(g)$ symmetry of its own, since in $\mathcal{N} = 4$ supersymmetric gauge theory a hypermultiplet gets a superpotential contribution which breaks this second $SU(g)$ symmetry. This superpotential contribution is of the form $\sim Q\Phi\tilde{Q}$ where $Q, \tilde{Q}$ are the $\mathcal{N} = 2$ chiral multiplets of the hypermultiplet and $\Phi$ the $\mathcal{N} = 2$ chiral multiplet of the $\mathcal{N} = 4$ vector-multiplet in the gauge theory. Overall, we expect $SU(g) \times U(1) \sim U(g)$ global symmetry. However, since the conjugate representations are real (adjoint) they are equivalent, which enhances the global symmetry from $U(g)$ (of dimension $g^2$) to $USp(2g)$ (of dimension $g(2g+1)$). See proof in Appendix A of [48]. Each hypermultiplet can be assigned a global symmetry fugacity from the set $\mathbf{a} \equiv \{a_i\}_{i=1}^g$, and these should form characters of $USp(2g)$ in the index.

- The symmetry between the $T[SU(N)]$ legs is the discrete $\mathbb{Z}_s$ which acts on composite operators, unlike the continuous symmetry between the hypermultiplets which acts on single fields. We will not deal with the discrete symmetry because even though it will change the CM globally, it will not change its dimension which we are interested in.

- Remember each $T[SU(N)]$ leg has a topological global symmetry $SU(N)_J$ represented by the fugacities $\mathbf{b}$. Since we have $s$ legs there are $s$ copies of $SU(N)_J$ symmetry and $s$ sets of fugacities $\{\mathbf{b}_i\}_{i=1}^s = \left\{\left\{b_i^{(k)}\right\}_{k=1}^{N-1}\right\}_{i=1}^s$. There is also a $U(1)_t$ global symmetry due to the R-symmetry (described in section B.4) represented by the fugacity $t$.

- In total, the global symmetry of the SSQ $(s, g, N)$ theory is $SU(N)_J^s \times USp(2g) \times U(1)_t$. One of the tests that the index is calculated correctly is that all the fugacities do in fact form characters of the corresponding global symmetry groups.

The index of the $SU(N)$ adjoint hypermultiplet with an additional fugacity $a$ is given by (based on the character of the adjoint):

$$I_{\text{hyp,adj}}\left(\mathbf{z}^{(N)}, \mathbf{m}^{(N)}, a; q, t\right) = I_{\text{hyp}}(a, 0; q, t)^{N-1} \prod_{i \neq j=1}^{N} I_{\text{hyp}}\left(\frac{z_i^{(N)}}{z_j^{(N)}} a, m_i^{(N)} - m_j^{(N)}; q, t\right) \tag{3.26}$$

The index of the SSQ $(s, g, N)$ theory in full glory:

$$\begin{aligned}
I_{\text{SSQ}(s,gN)}\left(\mathbf{a}, \{\mathbf{b}_i\}_{i=1}^s; q, t\right) \equiv \sum_{\mathbf{m}^{(N)}} \oint \frac{d\mathbf{z}^{(N)}}{2\pi i \cdot \mathbf{z}^{(N)}} I_{V,\mathcal{N}=4}^{SU(N)}\left(\mathbf{z}^{(N)}, \mathbf{m}^{(N)}; q, t\right) \\
\cdot \prod_{i=1}^{g} I_{\text{hyp,adj}}\left(\mathbf{z}^{(N)}, \mathbf{m}^{(N)}, a_i; q, t\right) \\
\cdot \prod_{j=1}^{s} I_{T[SU(N)]}\left(\mathbf{z}^{(N)}, \mathbf{m}^{(N)}, \mathbf{b}_j; q, t\right)
\end{aligned} \tag{3.27}$$

The most complicated part of this index stems from the $T[SU(N)]$ index. For $s = 0$ the $T[SU(N)]$ index is unnecessary and the SSQ index simplifies significantly.

### 3.4.1 Calculation and conjecture for the index

We calculated the index of SSQ $(s, g, N)$ for the parameter space spanned $0 \leq s \leq 5$, $1 \leq g \leq 5$ and $2 \leq N \leq 4$ (also $N = 5, 6$ for fewer cases), and based on these results conjecture the form of the index for any $s, g$ to first order in $q$ (also for any $N$ in case $s = 0$). For $s > 0$ we used the $T[SU(N)]$ indices from the previous section given as expansions to first order in $q$. We remind that the case $N = 2, s = 0, g = 1$ is "bad" and the index does not converge in this method, as expected.

In this theory the $\mathbf{z}^{(N)}$ fugacities do not appear any more since they are integrated in the central node $SU(N)$ gauge group. The characters of the different global symmetry group factors $USp(2g)$ and $\left\{SU(N)_J^i\right\}_{i=1}^s$ will be indicated in the superscript of the character. The representations of $USp(2g)$ will be described in terms of Dynkin labels [51], similarly to the treatment for $SU(N)$ already described in the previous section.

We conjecture that the index as a function of $s, g, N$ in the expansion to first order in $q$ is

$$I_{\text{SSQ}}(s, g, N) = 1 + I_{q^{\frac{1}{2}}}(s, g, N) q^{\frac{1}{2}} + I_{q^{\frac{3}{4}}}(s, g, N) q^{\frac{3}{4}} + I_q(s, g, N) q \tag{3.28}$$

where the components are:

$$I_{q^{\frac{1}{2}}}(s,g,N) = t \cdot \chi_{\text{adj}}^{USp(2g)} + \frac{1}{t}\sum_{i=1}^{s} \chi_{\text{adj}}^{SU(N)_i} \tag{3.29}$$

$$I_{q^{\frac{3}{4}}}(s,g,N) = \frac{1}{t^{\frac{1}{2}}}\chi_{[1,0,\ldots,0]}^{USp(2g)} + t^{\frac{3}{2}}\left(\chi_{[0,0,1,0,\ldots,0]}^{USp(2g)} + \delta_{N\geq 3} \cdot \chi_{[3,0,\ldots,0]}^{USp(2g)} + s \cdot \chi_{[1,0,\ldots,0]}^{USp(2g)}\right) \tag{3.30}$$

$$\begin{aligned}I_q(s,g,N) = &\underbrace{-1 - \chi_{\text{adj}}^{USp(2g)} - \sum_{i=1}^{s} \chi_{\text{adj}}^{SU(N)_i}}_{\text{conserved currents}} \\ &+ \delta_{N\geq 3} \cdot \chi_{\text{adj}}^{USp(2g)} + \chi_{\text{adj}}^{USp(2g)} \sum_{i=1}^{s} \chi_{\text{adj}}^{SU(N)_i} \\ &+ \frac{1}{t^2}\left(1 + \sum_{i=1}^{s} A_i(N) + \sum_{i\neq j}^{s} \chi_{\text{adj}}^{SU(N)_i}\chi_{\text{adj}}^{SU(N)_j}\right) \\ &+ t^2\left(B(g,N) + s \cdot \chi_{[0,1,0,\ldots,0]}^{USp(2g)} + \delta_{N\geq 3} \cdot s \cdot \chi_{\text{adj}}^{USp(2g)} + \frac{s(s-1)}{2}\right)\end{aligned} \tag{3.31}$$

In the expression above we used the subscript $i$ to indicate the different $SU(N)_J$ symmetries associated with the $T[SU(N)]$ legs (using the $\{\mathbf{b}_i\}_{i=1}^{s}$ fugacities). The adjoint representation of the $USp(2g)$ symmetry is given in terms of Dynkin labels by $\chi_{\text{adj}}^{USp(2g)} = \chi_{[2,0,\ldots,0]}^{USp(2g)}$ [51], and $B(g,N)$ is a shorthand notation for several characters of $USp(2g)$:

$$B(g,N) \equiv \begin{cases} \chi_{[4,0,\ldots,0]}^{USp(2g)} + \chi_{[0,2,0,\ldots,0]}^{USp(2g)} & N=2 \\ 1 + \chi_{[4,0,\ldots,0]}^{USp(2g)} + 2\chi_{[0,2,0,\ldots,0]}^{USp(2g)} + 2\chi_{[0,1,0,\ldots,0]}^{USp(2g)} + \chi_{[1,0,1,\ldots,0]}^{USp(2g)} & N=3 \\ 1 + 2\chi_{[4,0,\ldots,0]}^{USp(2g)} + 2\chi_{[0,2,0,\ldots,0]}^{USp(2g)} + 2\chi_{[0,1,0,\ldots,0]}^{USp(2g)} + \chi_{[1,0,1,\ldots,0]}^{USp(2g)} & N=4,5,6 \end{cases} \tag{3.32}$$

This form for $N = 5, 6$ was calculated only for a few cases of $s, g$ (for $N = 5$ on $s = 0, 2 \leq g \leq 4$ and for $N = 6$ on $s = 0, g = 2$ only). Regardless, we might conjecture this form is true for all $s, g$, and that the form of $B(g, N)$ for $N > 6$ remains the same as for $N = 4, 5, 6$. At least for the simple case $s = 0$ (without legs) we have a closed form conjecture for the index for all $g, N$ (since the result is independent of $A(N)$ which we do not presently know for $N > 5$).

We also note that for $g = 1$ the index has many additional exceptional terms left out of the conjectured form above and we do not bother listing them all. For $g = 2$ there is a single additional exceptional term $\frac{q}{t^2}\delta_{N=2,s=0,g=2}$.

The calculations and subsequent fitting of the fugacities with characters of the global symmetry groups were performed in an accompanying Mathematica notebook [52][53].

## 3.5 Z-extremization

As already discussed in section 3.1, R-symmetry has some freedom in its definition to it and can sometimes be mixed with $U(1)$ global symmetries. In SCFTs this is no longer the case and the R-charges take specific values that are consistent with the superconformal symmetry. The method of Z-extremization allows to calculate these values for 3d theories with at least 4 supercharges ($\mathcal{N} \geq 2$) [32]. For 3d theories that are $\mathcal{N} > 2$ (such as the theories we consider in this thesis) the R-symmetry is the non-abelian $SO(\mathcal{N})$ (see section 2.2) and no mixing is possible with $U(1)$ global symmetries, so the UV R-charges have to coincide with the IR values. Therefore, in these cases using the method of Z-extremization is merely a validation.

The object we consider is the function $Z$, which is the partition function of the 3d theory calculated on the compact manifold $\mathbb{S}^3$ (unlike the 3d supersymmetric index, which is the partition function calculated on $\mathbb{S}^2 \times \mathbb{S}^1$). The function $Z$ for a gauge theory is given by [32]

$$Z(\{\Delta_i\}) \equiv \int_{-\infty}^{\infty} d\mathbf{u} \left( \prod_{\alpha \in \text{rt}(G)} \sinh(\pi \cdot \alpha(u)) \right) \prod_{R_i} \prod_{\mu \in \text{wt}(R_i)} e^{l(1-\Delta_i + i\mu(u))} \quad (3.33)$$

where:

1. $\mathbf{u}$ are $\text{rank}(G)$ variables associated with the gauge group $G$ and $\int_{-\infty}^{\infty} d\mathbf{u}$ is shorthand for an integration over all the variables $\prod_{k=1}^{\text{rank}(G)} \int_{-\infty}^{\infty} du_k$.

2. $\alpha$ are the roots of the gauge group, $\{R_i\}$ are the representations of the $\mathcal{N} = 2$ chiral multiplets in the theory, $\mu$ are the weights of the representations, and $\alpha(u), \mu(u)$ are defined similary to $\alpha(m), \mu(m)$ from section 3.2.1.

3. $\Delta_i$ is the R-charge of the $\mathcal{N} = 2$ chiral multiplet in representation $R_i$.

4. The function $l(z)$ is defined as

$$l(z) \equiv -z \ln\left(1 - e^{2\pi i z}\right) + \frac{i}{2}\left(\pi z^2 + \frac{1}{\pi}\right) \text{Li}_2(x) - \frac{i\pi}{12} \quad (3.34)$$

where $\text{Li}_2(x)$ is the polylogarithm function.

According to [32], the values of the R-charges $\{\Delta_i\}$ that extremize the absolute value of the $Z$ function ($\partial_\Delta |Z| = 0$) are the superconformal R-charges.

Considering $\mathcal{N} = 4$ theories, the basic building blocks are $\mathcal{N} = 4$ hypermultiplets and an $\mathcal{N} = 4$ vector multiplet. The superpotential contains a contribution of the form $\sim Q\Phi\tilde{Q}$ where $Q, \tilde{Q}$ are the $\mathcal{N} = 2$ chiral multiplets of the hypermultiplet and $\Phi$ the $\mathcal{N} = 2$ chiral multiplet of the $\mathcal{N} = 4$ vector-multiplet. As discussed in section 3.2.2 and Appendix B.4, an $\mathcal{N} = 4$ theory can be described as an $\mathcal{N} = 2$ theory (with an abelian $U(1)_R$ R-symmetry) that has an additional global symmetry $U(1)_t$, where the $U(1)_t$ charges are the same for both components $Q, \tilde{Q}$ of the hypermultiplet.

The canonical assignment of the R-charges of $Q, \tilde{Q}$ is $\frac{1}{2}$ (see Table B.5 in Appendix B.3), and since the total R-charge of the superpotential is 2 [2] it implies that the R-charge of $\Phi$ is 1. The $U(1)_t$ charge of $\Phi$ is assigned such that the superpotential term is neutral ($U(1)_t$ singlet). We can summarize the different charge assignments in the following table:

Table 3.1: $U(1)$ charges of the fields $Q, \Phi, \tilde{Q}$

|  | $Q$ | $\Phi$ | $\tilde{Q}$ |
|---|---|---|---|
| $U(1)_R$ | $\frac{1}{2}$ | 1 | $\frac{1}{2}$ |
| $U(1)_t$ | 1 | $-2$ | 1 |

Now we allow for a general mixing of the R-symmetry with the the $U(1)_t$ symmetry as $R' = R + \alpha q_t$, where $\alpha$ is a mixing coefficient. The R-charges are now:

Table 3.2: Mixed $U(1)_R$ charges of the fields $Q, \Phi, \tilde{Q}$

|  | $Q$ | $\Phi$ | $\tilde{Q}$ |
|---|---|---|---|
| $U(1)_{R'}$ | $\Delta_Q \equiv \frac{1}{2} + \alpha$ | $\Delta_\Phi \equiv 1 - 2\alpha$ | $\Delta_{\tilde{Q}} \equiv \frac{1}{2} + \alpha$ |

We can now test whether the "unmixed" situation we expect is consistent with the correct superconformal values for the R-charges, for any 3d $\mathcal{N} = 4$ theory we are interested in. The theories analyzed in this thesis have only a single $U(1)$ global symmetry which is $U(1)_t$, so the Z-extremization has to be performed with respect to a single mixing parameter $\alpha$.

Consider the SSQ theories with $s = 0, N = 2$ for different $g$. This means we consider an $SU(2)$ gauge theory with $g$ $\mathcal{N} = 4$ hypermultiplets that are adjoints of $SU(2)$. For more details on the construction of these theories see section 3.4. The function $Z$ is given by

$$Z = -\int_{-\infty}^{\infty} du \cdot \sinh^2(2\pi u) \qquad (3.35)$$
$$\cdot \exp\left[g \cdot \left(l_{\text{adj}}^{SU(2)}(\Delta_Q, u) + l_{\text{adj}}^{SU(2)}(\Delta_{\tilde{Q}}, u)\right) + l_{\text{adj}}^{SU(2)}(\Delta_\Phi, u)\right]$$

where

$$l_{\text{adj}}^{SU(2)}(\Delta, u) \equiv l(1 - \Delta) + l(1 - \Delta + 2iu) + l(1 - \Delta - 2iu) \qquad (3.36)$$

is the contribution of an $SU(2)$ adjoint $\mathcal{N}=2$ chiral multiplet (in accordance with the $SU(2)$ adjoint character $\chi_{\text{adj}}^{SU(2)}(z) = 1 + z^2 + z^{-2}$).

Note that in this case the contributions of $Q, \tilde{Q}$ to the $Z$ function are equivalent. We performed the extremization numerically in an accompanying Mathematica notebook [52][53] by plotting $|Z|$ as a function of $\alpha$. The results indicate that for $g = 2, 3, 4$ (and by conjecture for higher $g$ as well) the extrema is indeed at the expected unmixed point $\alpha = 0$. However, for $g = 1$ the integrand of the $Z$ function diverges and so no conclusion can be reached. In this case the theory is analyzed by considering a dual theory, which we mentioned in section 3.4.

For the $T[SU(2)]$ theory, we have a $U(1)$ gauge theory with a bifundamental hypermultiplet of the $U(1)$ gauge group and $SU(2)$ global symmetry group. For more details on the construction of this theory see section 3.3. The $Z$ function is given by

$$Z = \int_{-\infty}^{\infty} du \cdot \exp\left[l_{\text{fund}}^{SU(2)}\left(\Delta_Q, u, 0\right) + l_{\text{fund}}^{SU(2)}\left(\Delta_{\tilde{Q}}, -u, 0\right) + l_{\text{adj}}^{U(1)}\left(\Delta_\Phi\right)\right] \quad (3.37)$$

where $l_{\text{adj}}^{U(1)}(\Delta_\Phi) \equiv l(1 - \Delta_\Phi)$ is the contribution of the $U(1)$ adjoint $\mathcal{N}=2$ chiral multiplet, and

$$l_{\text{fund}}^{SU(2)}(\Delta, u_1, u_2) \equiv l\left(1 - \Delta + i(u_1 + u_2)\right) + l\left(1 - \Delta + i(u_1 - u_2)\right) \quad (3.38)$$

is the contribution of an $SU(2)$ fundamental $\mathcal{N}=2$ chiral multiplet (in accordance with the $SU(2)$ fundamental character $\chi_{\text{fund}}^{SU(2)}(z) = z + z^{-1}$), that is also $U(1)$ charged. The variable $u_1$ is associated with $U(1)$ and $u_2$ associated with $SU(2)$. The global symmetry is not relevant for $Z$-extremization, so we can plug $u_2 = 0$ in the $Z$ function. Once again we verified $|Z|$ is extremized for $\alpha = 0$.

Finally, we look at the SSQ theories with $s = 1, N = 2$ for different $g$. In these theories we have $SU(2)$ and $U(1)$ gauge groups. The $Z$ function is given by:

$$\begin{aligned}Z = -\int_{-\infty}^{\infty} du_2 \cdot \sinh^2(2\pi u_2) \int_{-\infty}^{\infty} du_1 \\ \cdot \exp\left[l_{\text{fund}}^{SU(2)}\left(\Delta_Q, u_1, u_2\right) + l_{\text{fund}}^{SU(2)}\left(\Delta_{\tilde{Q}}, -u_1, u_2\right) + l_{\text{adj}}^{U(1)}(\Delta_\Phi)\right] \\ \cdot \exp\left[g \cdot \left(l_{\text{adj}}^{SU(2)}(\Delta_Q, u_2) + l_{\text{adj}}^{SU(2)}(\Delta_{\tilde{Q}}, u_2)\right) + l_{\text{adj}}^{SU(2)}(\Delta_\Phi, u_2)\right]\end{aligned} \quad (3.39)$$

We again verified $|Z|$ is extremized for $\alpha = 0$ for $1 \leq g \leq 4$, and we conjecture it remains so for higher $g$ as well.

# Chapter 4

# Conformal Manifold

## 4.1 Definition

As discussed in the introduction (chapter 1), the CM of a SCFT is given by the quotient of the space of marginal couplings and the complexified global symmetry group, and the DCM is the dimension of this quotient space. The mathematical object that will allow us to calculate it is the Hilbert series.

First, let us define the plethystic exponential (PE) of some multi-variable function $g(\mathbf{t})$ (that vanishes at the origin $g(\mathbf{0}) = 0$) as

$$\text{PE}\left[g\left(\mathbf{t}\right)\right] \equiv \exp\left(\sum_{n=1}^{\infty} \frac{g\left(\mathbf{t}^n\right)}{n}\right) \tag{4.1}$$

where by $\mathbf{t}^n$ we mean separately taking each of the variables of the function $g$ to the power $n$ [54][55][15]. Next, suppose we have an operator $\Phi$ in a representation $R$ of some group $G$ (that for example can represent a gauge group or a global symmetry group). Let us choose the function $g(\mathbf{t}, r) = r \cdot \chi_R(\mathbf{t})$ where $\chi_R(\mathbf{t})$ is the character of the representation $R$ (written in terms of the maximal torus variables $\mathbf{t}$) and $r$ is a free parameter (such that $|r| < 1$). The PE of this function gives

$$\text{PE}\left[r \cdot \chi_R\left(\mathbf{t}\right)\right] = \exp\left(\sum_{n=1}^{\infty} \frac{r^n \cdot \chi_R\left(\mathbf{t}^n\right)}{n}\right) = \sum_{n=0}^{\infty} r^n \cdot \chi_{\text{sym}^n(R)}\left(\mathbf{t}\right) \tag{4.2}$$

where $\chi_{\text{sym}^n(R)}(\mathbf{t})$ is the character of the symmetrized product of $n$ copies of the representation $R$ [56]. The PE performs the combinatorics involved with calculating the symmetrization of products of representations, meaning it is a generator for symmetrization [55][15]. If we project this object on the trivial representation, the result will be a series (called the Hilbert series) that counts the number of independent operators invariant under the group $G$ operation (group singlets) that can be formed using products of the operator $\Phi$. Singlets formed as products of $n$ operators are encoded in the $r^n$ coefficient of the series. This construction can be generalized to many operators $\{\Phi_i\}$ in different representations $\{R_i\}$ by plugging in $\sum_i \chi_{R_i}(\mathbf{t})$ to the PE [55][57]. The projection is done by integrating the maximal torus variables over the group $G$ with the Haar measure of the group [55][42]:

$$\mathcal{H}(r) \equiv \int_G \mathrm{d}\mu_G(\mathbf{t}) \, \mathrm{PE}\left[r \cdot \chi_R(\mathbf{t})\right] \tag{4.3}$$

See Appendix A.2 for more details on the Haar measure and its form for a general Lie group. We have now defined the Hilbert series. An important property of the Hilbert series is that it can be written as a rational function whose numerator and denominator are polynomials in $r$, and the power of the leading pole at $|r| = 1$ (degree of divergence) is the dimension of the manifold spanned by the operators [55].

We now apply this mathematical tool for our purpose. As defined in the introduction, the conformal manifold (CM) is spanned by the scalar operators formed using the exactly marginal couplings/operators. Following arguments based on the superconformal symmetry, the CM of a SCFT is given by the quotient of the space of marginal couplings/operators $\{\Phi_i^{\mathrm{marg}}\}$ by the complexified global symmetry group $G^{\mathbb{C}}$ [8]. Equivalently, we can replace $G^{\mathbb{C}}$ by $G$, given we form the scalar operators holomorphically in the marginal operators (meaning, taking each operator $\Phi$ without its complex conjugate $\bar{\Phi}$)[1] [8]. The Hilbert series does exactly that since it only counts singlets formed from products of the operators, without invoking their complex conjugates.

Using the supersymmetric index, in chapter 3 we deduced the marginal operators and the global symmetry of the SSQ theories. These operators come in different representations $\{R_i\}$ of the global symmetry group $G$. Therefore, plugging the sum of characters of these representations $\sum_i \chi_{R_i}^{\mathrm{marg}}(\mathbf{t})$ (without adding their complex conjugates as well) into the Hilbert series will perform the quotient we are looking for. Finally, the power of the leading pole of the Hilbert series (its degree of divergence) will capture the dimension of the conformal manifold (DCM).

The content of the rest of this chapter is as following:

- In section 4.2, we calculate the Hilbert series for a few simple examples to acquire some intuition for the procedure.
- In section 4.3, we define an algorithm to calculate the dimension from the Hilbert series in the general case.
- In section 4.4, we show that in some cases a simple formula can predict the dimension.
- In section 4.5, we apply the formula for the case of the SSQ theories and calculate their DCM, which is our main result.

---

[1] For a related derivation see Ch. 12.3 of [2].

## 4.2 Simple examples

A useful mathematical relation we will use multiple times in this section is

$$\text{PE}[x] = \exp\left(\sum_{n=1}^{\infty} \frac{x^n}{n}\right) = \exp\left(-\ln(1-x)\right) = \frac{1}{1-x} \quad (4.4)$$

for $|x| < 1$.

### 4.2.1 Trivial group

As a first simple example, take $N \in \mathbb{N}$ operators in the trivial representation ($\chi_e(\mathbf{z}) = 1$) of the trivial group $G = \{e\}$. The Hilbert series:

$$\begin{aligned}
\mathcal{H}(r) &= \text{PE}\left[r \cdot \sum_{i=1}^{N} \chi_e(\mathbf{z})\right] = \text{PE}[r \cdot N] = \exp\left(N \sum_{n=1}^{\infty} \frac{r^n}{n}\right) \\
&= \exp(-N\ln(1-r)) = \frac{1}{(1-r)^N} \\
&= \sum_{k=0}^{\infty} \binom{N+k-1}{k} r^k = 1 + Nr + \frac{N(N+1)}{2} r^2 + \frac{N(N+1)(N+2)}{3!} r^2 + \ldots
\end{aligned} \quad (4.5)$$

We have written down the series form as well as the rational function form. In this simple example, we have $N$ independent singlet operators and so the dimension of the manifold spanned by the operators is $N$ as expected.

### 4.2.2 $U(1)$ group

In the next example we take two operators oppositely charged under the $U(1)$ group. The characters of the operators are $\chi(z) = z^{\pm 1}$. The Haar measure for this group is:

$$\int_G \mathrm{d}\mu_G = \oint_{|z|=1} \frac{\mathrm{d}z}{2\pi i \cdot z} \quad (4.6)$$

The Hilbert series:

$$\begin{aligned}
\mathcal{H}(r) &= \int_G \mathrm{d}\mu_G \text{PE}\left[r\left(z + z^{-1}\right)\right] = \oint_{|z|=1} \frac{\mathrm{d}z}{2\pi i \cdot z} \text{PE}[r \cdot z] \text{PE}\left[r \cdot z^{-1}\right] \\
&= \oint_{|z|=1} \frac{\mathrm{d}z}{2\pi i \cdot z} \frac{1}{(1-rz)} \frac{1}{(1-rz^{-1})}
\end{aligned} \quad (4.7)$$

To calculate this contour integral we will use the Residue theorem, which states that the contour integral of a complex function $f(z)$ along a contour $\gamma$ given by

$$\oint_\gamma f(z)\, dz = 2\pi i \sum_i \text{Res}(f, z_i) \tag{4.8}$$

where $\{z_i\}$ are singular points of $f(z)$ enclosed inside the contour $\gamma$, and $\text{Res}(f, z_i)$ is the residue of $f$ at the point $z_i$. In our case we need to find the set of poles $\{z\}_{poles}$ of the integrand (zeros of the denominator) such that $|z| \leq 1$. There are 2 relevant poles at $z = 0, r$, and only the pole at $z = r$ contributes a non-zero value for the residue:

$$\mathcal{H}(r) = \oint_{|z|=1} \frac{dz}{2\pi i \cdot z} \frac{1}{(1 - r \cdot z)} \frac{1}{(1 - r \cdot z^{-1})} = \frac{1}{1 - r^2} = 1 + r^2 + r^4 + r^6 + \ldots \tag{4.9}$$

The dimension of the manifold spanned by the operators in this case is 1, since all contributions must come in pairs of the oppositely charged operators.

Note that if we take just one of these operators charged under $U(1)$, no singlet operator can be formed. Indeed, a direct calculation yields $\mathcal{H}(r) = 1$ which means the DCM is zero.

Generalizing to $N \in \mathbb{N}$ pairs of oppositely charged operators, the Hilbert series obtains the form

$$\mathcal{H}(r) = \oint_{|z|=1} \frac{dz}{2\pi i \cdot z} \frac{1}{(1 - r \cdot z)^N} \frac{1}{(1 - r \cdot z^{-1})^N} = \frac{P(r)}{(1 - r^2)^{2N-1}} \tag{4.10}$$

where $P(r)$ is some irreducible polynomial. We conjecture this based on the analytic solution for different values of $N$. Therefore, the dimension of the manifold spanned by the operators is $2N - 1$, which is the total number of operators minus one (the constraint of the $U(1)$ symmetry).

### 4.2.3 $SU(2)$ group

$SU(2)$ group is rank 1 and so has only one maximal torus variable. The Haar measure for this group is:

$$\int_G d\mu_G = \oint_{|z|=1} \frac{dz}{2\pi i \cdot z} \frac{1}{2} \left(1 - z^2\right) \left(1 - z^{-2}\right) \tag{4.11}$$

The characters of the fundamental and adjoint representations:

$$\begin{aligned} \chi_{\text{fund}}(z) &= z + z^{-1} \\ \chi_{\text{adj}}(z) &= 1 + z^2 + z^{-2} \end{aligned} \tag{4.12}$$

For a single $SU(2)$ fundamental operator $\Phi$, the Hilbert series (again using the

Residue theorem):

$$\mathcal{H}(r) = \oint_{|z|=1} \frac{dz}{2\pi i \cdot z} \frac{\frac{1}{2}(1-z^2)(1-z^{-2})}{(1-rz)(1-rz^{-1})} = 1 \quad (4.13)$$

No singlet can be formed which means the dimension of the manifold is zero. For a pair of fundamental and anti-fundamental operators $\Phi, \tilde{\Phi}$ (which for $SU(2)$ is the same as two fundamentals because all representations are real):

$$\mathcal{H}(r) = \oint_{|z|=1} \frac{dz}{2\pi i \cdot z} \frac{\frac{1}{2}(1-z^2)(1-z^{-2})}{(1-rz)^2(1-rz^{-1})^2} = \frac{1}{1-r^2} \quad (4.14)$$

The dimension is 1, corresponding to the singlet operator $\mathrm{tr}\left(\Phi\tilde{\Phi}\right)$. For an adjoint operator $\Phi$:

$$\mathcal{H}(r) = \oint_{|z|=1} \frac{dz}{2\pi i \cdot z} \frac{\frac{1}{2}(1-z^2)(1-z^{-2})}{(1-r)(1-rz^2)(1-rz^{-2})} = \frac{1}{1-r^2} \quad (4.15)$$

The dimension is again 1, corresponding to the singlet operator $\mathrm{tr}(\Phi^2)$. For a pair of adjoint operators $\Phi, \tilde{\Phi}$:

$$\mathcal{H}(r) = \oint_{|z|=1} \frac{dz}{2\pi i \cdot z} \frac{\frac{1}{2}(1-z^2)(1-z^{-2})}{(1-r)^2(1-rz^2)^2(1-rz^{-2})^2} = \frac{1}{(1-r^2)^3} \quad (4.16)$$

The dimension has increased to 3, corresponding to the singlet operators $\mathrm{tr}(\Phi^2)$, $\mathrm{tr}\left(\tilde{\Phi}^2\right)$ and $\mathrm{tr}\left(\Phi\tilde{\Phi}\right)$.

## 4.3 Algorithm for the general case

The character of a representation $R$ of group $G$ is given by

$$\chi_R(\mathbf{z}) = \sum_{\mu \in \mathrm{wt}(R)} z^\mu = \sum_{i=1}^{\dim(R)} \prod_{j=1}^{\mathrm{rank}(G)} z_j^{\mu_{i,j}} \quad (4.17)$$

where $\mu_{i,j}$ are the components of the weights of the representation (see Appendix A.1).

The Hilbert series of a representation $R$ can therefore be written as:

$$\begin{aligned}
\mathcal{H}(r) &= \int_G d\mu_G(\mathbf{z}) \, \mathrm{PE}\left[r \cdot \chi_R(\mathbf{z})\right] \\
&= \int_G d\mu_G(\mathbf{z}) \, \mathrm{PE}\left[r \cdot \sum_{i=1}^{\dim(R)} \prod_{j=1}^{\mathrm{rank}(G)} z_j^{\mu_{i,j}(R)}\right] \\
&= \int_G d\mu_G(\mathbf{z}) \prod_{i=1}^{\dim(R)} \frac{1}{\left(1 - r \prod_{j=1}^{\mathrm{rank}(G)} z_j^{\mu_{i,j}(R)}\right)}
\end{aligned} \quad (4.18)$$

For a set of representations $\{R_k\}_{k=1}^{N_R}$ the Hilbert series is given by:

$$\begin{aligned} \mathcal{H}(r) &= \int_G \mathrm{d}\mu_G(\mathbf{z}) \, \mathrm{PE}\left[r \cdot \sum_{k=1}^{N_R} \chi_R(\mathbf{z})\right] \\ &= \int_G \mathrm{d}\mu_G(\mathbf{z}) \prod_{k=1}^{N_R} \prod_{i=1}^{\dim(R_k)} \frac{1}{\left(1 - r \prod_{j=1}^{\mathrm{rank}(G)} z_j^{\mu_{i,j}(R_k)}\right)} \end{aligned} \quad (4.19)$$

In the simple rank 1 examples in the previous section, the calculation consisted of finding the poles of the integrand and using the Residue theorem. For a rank$>1$ group we have multiple integrations on the set of maximal torus variables $\{z_i\}$.

A recursive algorithm for the integration over variable $z_i$ (starting at level $i=1$):

1. Given the integrand, decompose denominator to monomials if possible (terms enclosed in parenthesis).

2. Calculate the zeros of all moniminals and collect them as candidates for poles $\{z_i\}_{\mathrm{poles}}$ (each pole can be a function of $r$ and $\{z_j\}$ for $i+1 \leq j \leq \mathrm{rank}(G)$, in total $1 + \mathrm{rank}(G) - i$ symbolic variables).

3. Collect set of unique pole candidates $\{z_i\}_{\mathrm{poles}}$ and throw out those that are outside of the unit circle $|z_i| > 1$ (dictated by the power of $r$, note that all the other variables $z_j$ are on the unit circle itself in the integration $|z_j| = 1$).

4. Calculate the residue for each pole and get the set $\{\mathrm{res}_i\}$. Not all the candidates are necessarily true poles, they might be singularities or regular points of the integrand yet we collect all of them for the residue calculations. If a candidate point is not a singularity at all, the residue will simply give zero. If a certain expression has no poles at all, it means the contour integral necessarily equals zero.

5. Repeat the algorithm for integration over the $z_{i+1}$ variable. This is basically the straightforward way to apply the Residue theorem for a multi-variable contour integral.

In Appendix C.2 we discuss the Mathematica [52] implementation of this algorithm.

## 4.4 Formula for the DCM of large representations

We applied the algorithm described in the previous section and calculated the dimension of the manifold for different representations of different Lie groups and summarised the results in Appendix C.3. In those calculations we found that for large enough (compared to the adjoint) generic representations the dimension of the manifold is given by the formula

$$\text{DM} = \dim(\text{reps}) - \dim(G) = \sum_i \dim(R_i) - \dim(\text{adj}). \tag{4.20}$$

If the group contains a $U(1)$ factor, then the formula breaks down even for large representations in case no representation is charged at all under the $U(1)$ or in the case the representations are charged in such a way that no $U(1)$ singlets can be formed at all. For a more elaborate discussion see Appendix C.3. The formula is not guaranteed to hold for any case we did not calculate directly, but we will assume that it does hold in the cases we analyze in the thesis.

## 4.5 DCM of the SSQ theories

We now move on to calculate the DCM for the SSQ theories, whose marginal operators and global symmetry were calculated and discussed in chapter 3.4. As explained in the introduction to this chapter, to calculate the DCM we need to calculate the dimension of the manifold spanned by the representations of the marginal operators using the Hilbert series. Since the SSQ theories contain many marginal operators the explicit calculation of the DCM using the algorithm of section 4.3 is numerically prohibitive. However, exactly because of that we are allowed to use the formula from section 4.4 applicable to large representations. The caveats to the formula do not apply in the SSQ theories so it is safe to use. We remind that the supersymmetric index contribution in power $q$ takes the following form:

$$I_q = \sum_i \chi_{R_i(G)}^{\text{marg}} - \chi_{\text{adj}(G)} \tag{4.21}$$

Therefore, according to the formula, the DCM can be written simply as $\dim(I_q)$, which means taking the $q$ coefficient of the index with all the global symmetry fugacities and the $t$ fugacity fugacities to one $I_q|_{\text{fug}\to 1}$ such that only the dimensions of the representations remain.

Since the SSQ theories are characterized by the integer parameters $s, g, N$, so will the DCM. Therefore, we first need to write down the dimensions of all representations

involved. For $USp(2g)$:

$$\begin{aligned}
\dim\left(\chi_{[1,0,\ldots,0]}^{USp(2g)}\right) &= 2g \\
\dim\left(\chi_{[2,0,\ldots,0]}^{USp(2g)}\right) &= g(2g+1) \\
\dim\left(\chi_{[0,1,\ldots,0]}^{USp(2g)}\right) &= (g-1)(2g+1) \\
\dim\left(\chi_{[0,2,\ldots,0]}^{USp(2g)}\right) &= \frac{1}{3}g\left(4g^3 - 7g + 3\right) \\
\dim\left(\chi_{[3,0,\ldots,0]}^{USp(2g)}\right) &= \frac{2}{3}g(g+1)(2g+1) \\
\dim\left(\chi_{[4,0,\ldots,0]}^{USp(2g)}\right) &= \frac{1}{6}g(g+1)(2g+1)(2g+3) \\
\dim\left(\chi_{[0,0,1,\ldots,0]}^{USp(2g)}\right) &= \frac{2}{3}(g-2)g(2g+1) \\
\dim\left(\chi_{[1,0,1,\ldots,0]}^{USp(2g)}\right) &= \frac{1}{2}(g-2)(g+1)(2g-1)(2g+1)
\end{aligned} \quad (4.22)$$

These forms were attained by fitting a polynominal in $g$ for the dimensions of the represetations for different $g$. Note these numbers all turn out to be integers. For the reps of $SU(N)$ for $2 \leq N \leq 5$ the relevant dimensions are:

$$\dim(A(N)) = \begin{cases} 5 & N=2 \\ 8+27 & N=3 \\ 15+20+84 & N=4 \\ 24+75+200 & N=5 \end{cases} = \begin{cases} 5 & N=2 \\ 35 & N=3 \\ 119 & N=4 \\ 299 & N=5 \end{cases} \quad (4.23)$$

These cases can be fit with a polynomial in $N$ in the form $\dim(A(N)) = 7N^3 - 36N^2 + 77N - 61$. The extrapolation to $N > 5$ is not to be trusted without explicit calculation. The dimension of the adjoint representation for any $N$ is $\dim\left(\chi_{\mathrm{adj}}^{SU(N)}\right) = N^2 - 1$. We can now plug these numbers into the the $q$ coefficient of the index $I_{\mathrm{SSQ}}(s,g,N)$ (Eq. 3.28) for $2 \leq N \leq 5$ (neglecting exceptional terms for low values of $g$) which gives the DCM:

$$\begin{aligned}
\mathrm{DCM}(s,g,N=2) &= \frac{1}{6}\left(12g^4 + 12g^3 + 48g^2s - 15g^2 + 12gs + 3g + 30s^2 - 24s\right) \\
\mathrm{DCM}(s,g,N=3) &= \frac{1}{6}\left(32g^4 + 120g^2s - 20g^2 + 48gs + 6g + 195s^2 - 39s\right) \\
\mathrm{DCM}(s,g,N=4) &= \frac{1}{6}\left(36g^4 + 12g^3 + 204g^2s - 9g^2 + 90gs + 9g + 678s^2 - 60s\right) \\
\mathrm{DCM}(s,g,N=5) &= \frac{1}{6}\left(36g^4 + 12g^3 + 312g^2s - 9g^2 + 144gs + 9g + 1731s^2 - 87s\right)
\end{aligned} \quad (4.24)$$

As explained in section 3.4.1, for $s = 0$ the result we conjecture to hold for $N \geq 4$

is given by:

$$\text{DCM}(s=0, g, N \geq 4) = \frac{1}{2}g\left(12g^3 + 4g^2 - 3g + 3\right) \tag{4.25}$$

These are the main results of the thesis. Note that regardless of the fractional prefactor, the DCM is always an integer as required. We can see that for all $N$ the DCM scales as $\sim s^2$ and $\sim g^4$ at the highest powers. This is very large compared to the 4d class $\mathcal{S}$ theories, where the DCM was calculated to scale linear both in $s$ and $g$. In 4d, the DCM was interpreted as a linear combination of geometric invariants of the Riemann surface $C_{g,s}$ the theory is defined with [9]. It is tempting to contemplate that the DCM of the 3d class $\mathfrak{s}$ theories (which are dual to the SSQ theories) could be written in terms of some geometric invariants of the Riemann surface times a circle $C_{g,s} \times \mathbb{S}^1$ the theory is defined with. We leave this as an open question for future research.

We note that the polynomial fitting of the dimensions of $USp(2g)$ representations and the resulting polynomial expression for the DCM as a function of $s, g$ was calculated in an accompanying Mathematica notebook [52][53]

# Chapter 5

# Conclusion

The strongly coupled regime of QFTs is of great interest both in high-energy and condenesd matter physics. Such regime can occur at the IR/UV fixed points of QFTs where the symmetry is enhanced to conformal symmetry. For supersymmetric QFTs which enjoy supersymmetry, the fixed points have superconformal symmetry. The conformal manifold (CM) is an interesting property of such fixed points, which is the space of all theories that originate as deformations of the original theory, yet retain the conformal symmetry. The dimension of the conformal manifold (DCM) is the number of independent exactly marginal operators that span the CM. In recent years a tool has been developed in the theoretical high-energy physics community that allows to calculate this property for supersymmetric theories, called the supersymmetric index. In chapter 3 we described the supersymmtric index and calculated the supersymmetric index for a class of 3d $\mathcal{N} = 4$ supersymmetric theories called the star-shaped-quiver theories, more elaboratly discussed in the introduction (chapter 1) and in chapter 3. It is possible to extract the exactly marginal operators and the global symmetry in the IR fixed point directly from the supersymmetric index, as explained in the introduction. The DCM can be calculated as the dimension of the quotient of the space of marginal couplings and the complexified global symmetry group. This problem now becomes group-theoretic, and can be approached using the Hilbert series which performs the above mentioned quotient. The numeric evaluation of the DCM using the Hilbert series becomes increasingly difficult for larger representations, however once the representations become large enough compared to the adjoint the DCM starts following a simple rule (with a few caveats), as we calculated and discussed in chapter 4. We assume this rule holds for general large representations, which should apply for the SSQ theories.

Our main result is the calculation of the DCM of the 3d SSQ theories, which are dual to the 3d class $\mathfrak{s}$ theories. We have found that the DCM scales at the highest power as $\sim g^4$ and $\sim s^2$. This is a large scaling compared to the DCM of other related theories, the 4d class $\mathcal{S}$ theories, where the scaling was found to be linear in both $g$ and $s$. The DCM of these 4d theories was interpreted as a linear combination of some geometric invariants of the Riemann surface $C_{g,s}$ the theory is defined with [9].

This is an indication that the DCM of the 3d class $\mathfrak{s}$ theories might also be able to be interpreted using the geometric invariants of the Riemann surface times a circle $C_{g,s} \times \mathbb{S}^1$ the theory is defined with. However, we have not been able to find such an interpretation so far, and this remains an intriguing open question for further research. Another interesting prospect is to calculate the DCM of other classes of 3d $\mathcal{N} = 4$ theories and see if the large DCM we found for the SSQ is a generic feature of 3d $\mathcal{N} = 4$ theories.

# Appendix A

# Lie Groups

## A.1 Characters

The character $\chi_R$ of a representation $R$ (of a group $G$) is defined as the trace of the representation matrix $\chi_R \equiv \text{tr}(D_R)$. For Lie groups, the representation matrices can be written as $D_R = e^{i \sum_i \theta_i T_i}$ where $\{T_i\}_{i=1}^{\dim(G)}$ are the $\dim(G)$ generators of the Lie algebra $\mathfrak{g}$ (corresponding to the Lie group $G$). Due to the trace in the definition of the character, only the mutually-commuting generators are necessary to define the character. The Cartan subalgebra $\mathfrak{h}$ is the largest subset of mutually-commuting generators, defined to be of size $\text{rank}(G)$ [42]. The Lie group $H$ corresponding to the Lie algebra $\mathfrak{h}$ is the largest abelian subgroup of $G$, called the maximal torus $T = U(1)^{\text{rank}(G)}$ [58][59]. For a connected compact Lie group $G$, the character can be written in terms of the maximal torus variables $\{z_i\}_{i=1}^{\text{rank}(G)}$ (parameters defined on the unit circle $|z_i| = 1$), in the form

$$\chi_R = \sum_{\mu \in \text{wt}(R)} \mathbf{z}^\mu = \sum_{i=1}^{\dim(R)} \prod_{j=1}^{\text{rank}(G)} z_j^{\mu_{i,j}} \quad (A.1)$$

where $\{\mu\}_{i=1}^{\dim(R)}$ are the weights of the representation [60][61]. Each weight is vector in $\text{rank}(G)$ dimensional space $\mu = \left(\mu_1, ..., \mu_{\text{rank}(G)}\right)$. $\mathbf{z}^\mu$ is a shorthand notation for $\prod_{j=1}^{\text{rank}(G)} z_j^{\mu_{i,j}}$ where $\mu_{i,j}$ is the $j$-th component of the $i$-th weight vector. By definition, after setting all variables to one $z_i = 1$ the character gives the dimension of the representation (size of the weight system).

The Mathematica [52] package LieART [51] is a very useful tool for dealing with representations of Lie groups, except it does not generate the character formulas in terms of the maximal torus variables. Using the WeightSystem command we can generate the weights $\{\mu\}$ of any representation $R$ charactereized by its Dynkin labels [51][57], and then plug them into Eq. A.1 to get the character form. We implemented this in the accompanying Mathematica package used for the thesis [53].

## A.2 Haar measures

Characters of irreducible representations are orthonormal with respect to integration over the group

$$\int_G d\mu_G \chi^*_{R_1} \chi_{R_2} = \delta_{R_1, R_2} \tag{A.2}$$

where $d\mu_G$ is the invariant measure (called the Haar measure) of the group [61][42]. The Haar measure of a Lie group $G$ is given by

$$\begin{aligned}\int_G d\mu_G &= \frac{1}{|W|} \left( \prod_{i=1}^{\text{rank}(G)} \oint_{|z_i|=1} \frac{dz_i}{2\pi i \cdot z_i} \right) \prod_{\alpha \in \text{rt}(G)} (1 - \mathbf{z}^\alpha) \\ &= \frac{1}{|W|} \oint \frac{d\mathbf{z}}{2\pi i \cdot \mathbf{z}} \prod_{\alpha \in \text{rt}(G)} (1 - \mathbf{z}^\alpha)\end{aligned} \tag{A.3}$$

where:

- $|W|$ is the order of the Weyl group $W$ of $G$ [61][62][63].
- $\{z_i\}_{i=1}^{\text{rank}(G)}$ are the maximal torus variables (defined in the previous section).
- The integration for each variable is over the unit circle $\int_0^{2\pi} \frac{d\theta}{2\pi i} = \oint_{|z|=1} \frac{dz}{2\pi i \cdot z}$ (using the change of variables $z = e^{i\theta}$, $\frac{dz}{iz} = d\theta$). In the second line we defined a shorthand notation for the integration over all variables $\oint \frac{d\mathbf{z}}{2\pi i \cdot \mathbf{z}}$.
- $\{\alpha\}_{i=1}^{\text{rank}(G)}$ are the roots of $G$, where each root is a vector $\alpha = \left(\alpha_1, ..., \alpha_{\text{rank}(G)}\right)$ in rank $(G)$ dimensional space. $\mathbf{z}^\alpha$ is a shorthand notation for $\prod_{i=1}^{\text{rank}(G)} z_j^{\alpha_{i,j}}$ where $\alpha_{i,j}$ is the $j$-th component of the $i$-th root vector (similar to the definitions for weights in the previous section).

The dimension of the Weyl groups for the different classes of compact Lie groups (for $n \in \mathbb{N}$) [61][62][63]:

$$\begin{aligned}|W(A_n)| &= |W(SU(n+1))| = (n+1)! \\ |W(B_n)| &= |W(SO(2n+1))| = 2^n n! \\ |W(C_n)| &= |W(Sp(2n))| = |W(B_n)| \\ |W(D_n)| &= |W(SO(2n))| = 2^{n-1} n! \\ |W(E_6)| &= 72 \cdot 6! \\ |W(E_7)| &= 72 \cdot 8! \\ |W(E_8)| &= 192 \cdot 10! \\ |W(F_4)| &= 1152 \\ |W(G_2)| &= 12\end{aligned} \tag{A.4}$$

For $G = U(1)$ there are no roots and the $|W| = 1$ so:

$$\int_G \mathrm{d}\mu_G = \oint_{|z|=1} \frac{\mathrm{d}z}{2\pi i \cdot z} \tag{A.5}$$

There is also a shorter version (without the $|W|$ normalization) for the Haar measure given by

$$\int_G \mathrm{d}\mu_G = \oint \frac{\mathrm{d}\mathbf{z}}{2\pi i \cdot \mathbf{z}} \prod_{\alpha \in \mathrm{rt}_+(G)} (1 - \mathbf{z}^\alpha) \tag{A.6}$$

where $\alpha$ are only the positive roots. This simplified form can speed up computations, especially for higher rank groups [61]. The Haar measures $\mathrm{d}\mu_G(\mathbf{t})$ for the classical Lie groups are conveniently summarized in [61].

The roots (or positive roots) of different Lie groups can be generated with the LieART package [51] using the RootSystem (or PositiveRoots) command. Note the roots are non-zero vectors, and the RootSystem command also gives the Cartan generators (rank $(G)$ zero vectors), which should be removed to get the correct set of roots. Also note the root system is the same as the weight system of the adjoint representation, with the zero weights omitted [61]. In the accompanying Mathematica package used for the thesis [53] we implemented the Haar measure based on the LieART commands (both "long" and "short" versions).

## A.3 Switching the basis

In the previous sections, we defined the characters using the weight system of a represention, and the Haar measure using the root system of the group. However, the forms we obtain do not necessarily coincide with the standard forms for the characters and Haar measures for different Lie groups [64][65][57]. The reason is that the definition of the maximal torus is not unique. Therefore, we shall perform a change of variables that will mix the variables $\{z_i\}$ among themselves such that we end up with the desired forms.

This mixing will be performed using a linear transformation. We will build an invertible matrix that will do the job. The weight/root vectors are rank $(G)$ dimensional. We take rank $(G)$ weight vectors (of the fundamental or adjoint representation) and stack them up to form a square matrix $A$ (choose independent weight vectors so $A$ will be invertible). Then choose a few vectors we would like these to be transformed into to get the correct form of the character, and stack them up as well to a matrix $B$ (again, choose independent vectors). The linear transformation matrix $M$ is defined to transform each row in $A$ to a row in $B$, meaning $M \cdot A^T = B$. Therefore, $M = B \cdot \left(A^T\right)^{-1}$. Note the choices made are not unique, the vectors and mapping between them can be chosen in many ways. We test the matrix $M$ by checking that the characters after the transformation (for the representations we have formulas for) are of the correct form.

For the classical Lie groups we can do this for a few examples and deduce the general rule for the form of $M$ for a general rank group. In an accompanying Mathematica notebook [53] we show these calculations explicitly. The results of these calculations follow. For $SU(N)$ and $Sp(2N)$, $N \geq 2$ [64]:

$$M = \begin{pmatrix} 1 & 1 & \cdots & 1 \\ & 1 & \cdots & 1 \\ & & \ddots & \vdots \\ & & & 1 \end{pmatrix} \tag{A.7}$$

For $SO(2N+1)$, $N \geq 3$ [65]:

$$M = \begin{pmatrix} 1 & 1 & \cdots & 1 & 1 & \frac{1}{2} \\ & 1 & \cdots & 1 & 1 & \frac{1}{2} \\ & & \ddots & \vdots & \vdots & \vdots \\ & & & 1 & \frac{1}{2} \\ & & & & & \frac{1}{2} \end{pmatrix} \tag{A.8}$$

For $SO(2N)$, $N \geq 4$ [65]:

$$M = \begin{pmatrix} 1 & 1 & \cdots & 1 & 1 & \frac{1}{2} & \frac{1}{2} \\ & 1 & \cdots & 1 & 1 & \frac{1}{2} & \frac{1}{2} \\ & & \ddots & \vdots & \vdots & \vdots & \vdots \\ & & & 1 & \frac{1}{2} & \frac{1}{2} \\ & & & & & \frac{1}{2} & \frac{1}{2} \\ & & & & & -\frac{1}{2} & \frac{1}{2} \end{pmatrix} \tag{A.9}$$

For $G_2$ [64]:

$$M = \begin{pmatrix} 1 & 1 \\ 0 & -1 \end{pmatrix} \tag{A.10}$$

Note the form of the $G_2$ Haar measure in [55] is without this transformation. The other exceptional Lie groups $F_4, E_6, E_7, E_8$ can in principle be treated the same way, but no exceptional groups were needed in this thesis so this was not performed.

### A.3.1 Switching the basis in the $T[SU(N)]$ theory

In chapter 3.3 we defined the $T[SU(N)]$ theories which are one of the key components of the SSQ theories. As explained, the $T[SU(N)]$ theory index depends on different fugacities: $q, t$ associated with the R-symmetry and Lonretz symmetry, $\mathbf{z}^{(N)}, \mathbf{m}^{(N)}$ the fugacities and fluxes of the $SU(N)$ global flavor symmetry, and $\mathbf{b}$ the fugacities associated with the topological global symmetries $U(1)_J$ from each abelian gauge factor ($N-1$ of these in total). The latter symmetry is expected to enhance from $U(1)^{N-1}$ to $SU(N)$ [50][11][41]. This means that in the index expression the fugacities **b** should form characters of $SU(N)$. For example, for $N=2$ there is a single fugacity $b$ which should form characters of $SU(2)$.

However, as we discussed in the previous section, the form of the characters might not coincide with the standard form of the character of $SU(N)$ (unlike the $\mathbf{z}^{(N)}$ fugacities which do this correctly by construction). Therefore, we will mix the **b** fugacities such that we arrive at the desired form. We do this by finding an appropriate weights transformation matrix $M$ of size $(N-1) \times (N-1)$. In the previous section the mapping was found for a specific representation (fundamental or adjoint) where we knew the initial (LieART generated) form and final (standard) form. Here the situation is more difficult because we do not know in advance which $SU(N)$ representations contribute to the index and exactly how.

The solution was found in the following way: First we calculated the index of the $T[SU(N)]$ theory in the zero global flux sector $\mathbf{m}^{(N)}=0$. Upon inspection we noticed that the coefficient of the index in power $q^{\frac{1}{2}}t$ depends only on the $\mathbf{z}^{(N)}$ fugacities which form the adjoint character of $SU(N)$ (tested for $2 \leq N \leq 5$). In power $q^{\frac{1}{2}}t^{-1}$ the coefficient depends only on the **b** fugacities, and at equal number of terms (size of the contributing representations) as the $q^{\frac{1}{2}}t$ coefficient. Therefore, we conjecture a symmetry between the $q^{\frac{1}{2}}t^{\pm 1}$ terms, meaning the **b** fugacities also contribute as the adjoint representation of $SU(N)_J$, and we search for the transformation matrix that will make the **b** fugacities form the adjoint character as well. This can be done by a bruteforce trial and error method for every $N$.

For $N=2$ there is only one maximal torus variable $b$, and for it to form standard characters of $SU(2)$ we need to rescale it $b \to b^2$ (equivalent to the $1 \times 1$ transformation matrix $M_{N=2}=(2)$), which is the normalization that appears in [11]. For $3 \leq N \leq 5$ we define the transformation matrix

$$M_{3 \leq N \leq 5} = \begin{pmatrix} 1 & -1 & 0 & 0 \\ 1 & 2 & 1 & 1 \\ -1 & -1 & -2 & -1 \\ 0 & 0 & 1 & -1 \end{pmatrix} \tag{A.11}$$

For $N=5$ there are 4 maximal torus variables and we use all the matrix, and for $N=3,4$ we take only the upper left part of the matrix. For example, for $N=3$:

$$M_{N=3} = \begin{pmatrix} 1 & -1 \\ 1 & 2 \end{pmatrix} \tag{A.12}$$

Note the these matrices are not unique. The full form of the supersymmetric indices of $T[SU(N)]$ using these transformation matrices appear in chapter 3.3.

## A.4 Description of a representation using a subgroup

As discussed in the section A.1, a character of some representation of a Lie group $G$ can be written in terms of rank $(G)$ maximal torus variables. If we choose to write the character in terms of the maximal torus variables of a subgroup of $G$ that has the same rank, no information is lost.

As an example, consider the character of the adjoint representation of $SU(2)$, written in terms of the maximal torus variable $a$:

$$\chi_{\mathbf{3}}^{SU(2)} = 1 + a^2 + a^{-2} \tag{A.13}$$

The character of the adjoint of any Lie group $G$ always has rank $(G)$ singlets (zero weights), which is one in this case. The only subgroup of $SU(2)$ is $U(1)$ [51]. We can think of $a$ as the maximal torus variable of $U(1)$, and then the character above decomposes to 3 operators charged with $U(1)$. The adjoint of a group always decomposes to the adjoint of the subgroup plus some more representations, and in this case the singlet of $SU(2)$ is the adjoint of $U(1)$.

Now consider the group $SO(4) \sim SU(2)_a \times SU(2)_b$, its adjoint is of dimension 6 since $\dim(SO(n)) = \frac{n(n-1)}{2}$. Since the adjoint of $SU(2)$ is **3**, the adjoint of $SU(2)_a \times SU(2)_b$ is $\mathbf{3}_a + \mathbf{3}_b$. The character is accordingly (written in terms of the maximal torus variables $a, b$):

$$\chi_{\text{adj}} = 1 + a^2 + a^{-2} + 1 + b^2 + b^{-2} \tag{A.14}$$

Define the generators of the $SU(2)$ factors as $J_a, J_b$. If we were to change variables and look at combinations of the generators we would lose the apparent $SU(2)$ symmetries (unless we just swap $J_a \leftrightarrow J_b$). For example, using the generators $J_a \pm J_b$ (motivated in chapter 3.2.2) means changing variables $a = uv, b = u/v$ (or using the weights transformation matrix $M = \begin{pmatrix} 1 & 1 \\ 1 & -1 \end{pmatrix}$ following the notation of the previous section) the character transforms to:

$$\chi_{\text{adj}} = 1 + (uv)^2 + (uv)^{-2} + 1 + \left(\frac{u}{v}\right)^2 + \left(\frac{u}{v}\right)^{-2} \tag{A.15}$$

We still have the two singlets since we are considering the adjoint of a rank 2 group. However, the $SU(2)$ symmetries are no longer apparent. Meaning, we are describing the representation in terms of the $U(1) \times U(1)$ subgroup (which has the same rank).

For decompositions of different representations of different Lie groups to their subgroups, see Appendix A.3 in [51].

# Appendix B

# Supersymmetric Index as Trace

As discussed in chapter 3, the supersymmetric index can be defined as a partition function or as a trace formula. Here we discuss the second approach, also called the superconformal index when calculated in the fixed points. This was discussed numerously in the literature for 4d [29][66][15] and 3d [67][68][[34][16] theories. Here, we aim to combine the derivations to a (hopefully) coherent story that will serve the goal of pedagogically explaining the 3d $\mathcal{N} = 4$ case relevant for the SSQ theories discussed in this thesis.

We begin with the definition of the superconformal index in general and then calculate the superconformal index building blocks first for the 4d $\mathcal{N} = 1$ case and then for the 3d $\mathcal{N} = 2, 4$ cases (in the zero monopole sectors) to the extent it is possible.

## B.1 Definition

In supersymmetry or superconformal symmetry, the anti-commutator of two different fermionic supersymmetry generators $\delta \equiv \left\{Q, Q^\dagger\right\}$ is some bosonic generator (momentum for supersymmetry and something more complex for superconformal symmetry). The Witten index $\text{Tr}_{\mathcal{H}}\left((-1)^F z^\delta\right)$ is a trace over all (supersymmetric) states in the Hilbert space, weighed by the fermion number of the state $(-1)^F$ and some exponent of $\delta$ (note that inside the trace when we write an operator we actually mean the charge of the state under the operator). $z$ is a new parameter called the fugacity. Alternatively, it is common to write $e^{-\nu\delta}$ where $\nu$ is a chemical potential. For states that are not annihilated by $\delta$, supersymmetry dictates that the state comes in equal number of bosons and fermions, and therefore their contribution to the index vanishes because of the $(-1)^F$. States annihilated by $\delta$ do contribute, but then the result is independent of $z$ [2][3]. In normal supersymmetry $\delta$ annihilates the zero-energy states, and therefore the index counts the number of bosonic minus fermionic vacua (ground states). In superconformal symmetry, the number of "vacua" (eigenstates of $\delta$ with zero eigenvalue) can be infinite, so the index is not well defined and needs some regularization, achieved using the following addition.

By weighing the trace with more operators $\{h_i\}$ (each with its own fugacity) that commute with $\delta$ (so their eigenstates are simultaneously diagonalizable), we can write a more general index $\text{Tr}_{\mathcal{H}}\left((-1)^F z^\delta \prod_i a_i^{h_i}\right)$. For non-zero fugacities $a_i \neq 1$ the index can be made finite for SCFTs. This index will encode properties of the conformal manifold, the object of interest in this thesis. Therefore, we need to identify the operators $\{h_i\}$ that commute with $\delta$ but are different from $\delta$ (their eigenvalues with the vacua of $\delta$ are non-zero), otherwise the index will independent of $\{a_i\}$.

In a superconformal algebra (of dimension $d$) the generators are (also see chapter 2.2):

- The dilatation operator $\Delta$.

- The rotation generators $M_{\mu\nu}$ (their number is the rank of rotation group $SO(d)$).

- Supersymmetry generators $Q_{i\alpha}$ (their number is the rank of the R-symmetry group $G_R$).

- There are also an equal number $\text{rank}(G_R)$ of special supersymmetry generators $S_{i\alpha}$, but they are not independent in the superconformal algebra and satisfy $S_{i\alpha} = Q_{i\alpha}^\dagger$. Also there's the momentum generator $P_\mu$ and the special conformal generators $K_\mu$ that are also not independent and given as anti-commutators of $\{Q, Q^\dagger\}$ and $\{S, S^\dagger\}$, respectively.

From the Bianchi identity it follows that both $Q$ and $S$ commute with $\delta$, so finding operators that commute with $\delta$ is the same as finding those that commute with $Q, S$. The number of bosonic generators we identified so far as independent is $n_{\text{gens}} = 1 + \text{rank}(SO(d)) + \text{rank}(G_R)$. $\delta$ itself is a one combination of these generators so we do not count it twice. The remaining $n_{\text{gens}} - 1$ generators do not necessarily commute with $\delta$, so we need the subspace that does. The subspace of linearly independent combinations of the remaining $n_{\text{gens}} - 1$ generators that commutes with $\delta$ is of size $n_{\text{gens}} - 2 = \lfloor d/2 \rfloor + \text{rank}(G_R) - 1$. Another way to come up with this number is to start with $n_{\text{gens}}$ generators, and applying the two restrictions that the generators commute with $\delta$ (or $Q, S$) and not be $\delta$.

On top of that, if there are additional global symmetries (that of course commute with all the spacetime related generators) they can also be added to the index. If the global symmetry has $m$ Cartan subalgebra generators $\{F_i\}$, they can be added to the index with more fugacity parameters $\prod_{i=1}^m \mu_i^{F_i}$ to give the full superconformal index:

$$I(\{a_i\}, \{\mu_j\}) \equiv \text{Tr}_{\mathcal{H}}\left((-1)^F z^\delta \prod_i a_i^{h_i} \prod_{j=1}^m \mu_j^{F_j}\right) \tag{B.1}$$

Examples of the fugacities possible for 3,4d:

- In 4d $\mathcal{N} = 1$ ($U(1)$ R-symmetry) we can write 2 generators, whose fugacities are usually called $p, q$. In 4d $\mathcal{N} = 2$ ($SU(2) \times U(1)$ R-symmetry) we can write an additional generator associated with the additional R-symmetry Cartan generator.

- In 3d $\mathcal{N} = 2$ ($SO(2) \sim U(1)$ R-symmetry) we can write 1 generator, whose fugacity is usually called $q$. In 3d $\mathcal{N} = 4$ ($SO(4) \sim SU(2) \times SU(2)$ R-symmetry) we can write an additional generator associated with the additional R-symmetry Cartan generator.

Formal constructions of the index for general R-symmetry were done for 4d [29] and 3,5,6d [67][68]. However, the general formulas are formal and complicated and below we will use formulas that are specific to the choice of $\mathcal{N}$ considered in each case.

We note that in 3d the $(-1)^F$ factor in the defnition is modified in the presence of magnetic monopole charges [69][47][11].

## B.2  4d $\mathcal{N} = 1$ index

The 4d $\mathcal{N} = 1$ multiplets are [2]:

- $\mathcal{N} = 1$ chiral multiplet, containing a scalar field $\phi$ and a spinor field $\psi_\pm$.

- $\mathcal{N} = 1$ vector multiplet, containing the vector field $A_\mu$ (or the gauge-invariant field strength tensor $F_{\mu\nu}$) and a spinor field $\lambda_\pm$.

The 4d $\mathcal{N} = 1$ right-handed index definition is [15]

$$I = \mathrm{Tr}_{\mathcal{H}} \left( (-1)^F e^{\delta} p^{j_1+j_2+\frac{1}{2}R} q^{j_2-j_1+\frac{1}{2}R} \mu^F \right)$$
$$\delta = \Delta - 2j_2 - \frac{3}{2}R \tag{B.2}$$

where $\Delta$ the dilatation generator, $j_1, j_2$ are the $SO(4) \sim SU(2) \times SU(2)$ Lorentz group Cartan subalgebra generators and $R$ the $U(1)$ R-symmetry generator.

In the following table we write down the available operators (that we call "letters"), their charges with respect to the different generators, and their contributions to the (single letter) index $i$ in case they satisfy $\delta = 0$, according to Eq. B.2. In the notes that follow we provide detailed explanations.

Table B.1: Charges and supersymmetric index contributions of different letters in 4d $\mathcal{N} = 1$

| Letter | Lorentz rep. | $\Delta$ | $j_1$ | $j_2$ | $R_{UV}$ | $R_{IR}$ | $\delta$ | $i$ |
|---|---|---|---|---|---|---|---|---|
| $Q_\pm$ | $(0,0)$ | $\frac{1}{2}$ | $0$ | $\pm\frac{1}{2}$ | $1$ | | | |
| $\phi$ | $(0,0)$ | $1$ | $0$ | $0$ | $\frac{2}{3}$ | $r$ | $0$ | $(pq)^{\frac{r}{2}}$ |
| $\bar\phi$ | $(0,0)$ | $1$ | $0$ | $0$ | $-\frac{2}{3}$ | $-r$ | $2$ | $0$ |
| $\psi_\pm$ | $\left(\frac{1}{2},0\right)$ | $\frac{3}{2}$ | $\pm\frac{1}{2}$ | $0$ | $-\frac{1}{3}$ | $r-1$ | $2^\pm$ | $0$ |
| $\bar\psi_\pm$ | $\left(0,\frac{1}{2}\right)$ | $\frac{3}{2}$ | $0$ | $\pm\frac{1}{2}$ | $\frac{1}{3}$ | $-r+1$ | $0^+, 2^-$ | $-(pq)^{\frac{2-r}{2}}, 0$ |
| $F_\pm$ | $(1,0)$ | $2$ | $\pm 1$ | $0$ | $0$ | $0$ | $2^\pm$ | $0$ |
| $F_0$ | $(1,0)$ | $2$ | $0$ | $0$ | $0$ | $0$ | $2$ | $0$ |
| $\bar F_\pm$ | $(0,1)$ | $2$ | $0$ | $\pm 1$ | $0$ | $0$ | $0^+, 4^-$ | $pq, 0$ |
| $\bar F_0$ | $(0,1)$ | $2$ | $0$ | $0$ | $0$ | $0$ | $2$ | $0$ |
| $\lambda_\pm$ | $\left(\frac{1}{2},0\right)$ | $\frac{3}{2}$ | $\pm\frac{1}{2}$ | $0$ | $1$ | $1$ | $0^\pm$ | $-p, -q$ |
| $\bar\lambda_\pm$ | $\left(0,\frac{1}{2}\right)$ | $\frac{3}{2}$ | $0$ | $\pm\frac{1}{2}$ | $-1$ | $-1$ | $\frac{5}{2}^+, \frac{7}{2}^-$ | $0$ |
| $\partial_{\pm +}$ | $\left(\frac{1}{2},\frac{1}{2}\right)$ | $1$ | $\pm\frac{1}{2}$ | $\frac{1}{2}$ | $0$ | $0$ | $0^\pm$ | $p, q$ |
| $\partial_{\pm -}$ | $\left(\frac{1}{2},\frac{1}{2}\right)$ | $1$ | $\pm\frac{1}{2}$ | $-\frac{1}{2}$ | $0$ | $0$ | $2^\pm$ | $0$ |

Notes:

- $\pm$ in the subscript indicates the spinor index, and $\delta$ for both options is calculated (the $\pm$ in the superscript indicates the spinor component for which it was calculated).

- The supersymmetry generator $Q_\pm$ is chosen to be the right-handed one [15], therefore its Lorentz charges are $j_1 = 0, j_2 = \pm\frac{1}{2}$. From the supersymmetry algebra $\Delta(Q) = \frac{1}{2}$ and $R(Q) = 1$ [2]. These charges are independent of the dimension.

- The dimension $\Delta$ values for the chiral superfield components $\phi, \psi$ are chosen from their canonical kinetic terms, $2[\phi] + 2 = d$ giving $[\phi] = \frac{d-2}{2}$ and $2[\psi] + 1 = d$ giving $[\psi] = \frac{d-1}{2}$.

- The $R_{UV}$ charge of the lowest component $\phi$ of the chiral superfield is assigned using the unitarity bound $\Delta = \frac{3}{2}|R_{UV}|$ in 4d. The $R_{UV}$ charge of $\psi$ can be

- assigned by looking at it as part of a chiral superfield, and so $R_{UV}(\psi) = R_{UV}(\phi) - 1 = -\frac{1}{3}$.

- The vector representation of $SO(4) \sim SU(2) \times SU(2)$ is $\mathbf{2} \times \mathbf{2}$. Its square decomposition is $(\mathbf{2} \times \mathbf{2})^2 = (\mathbf{1} \times \mathbf{1} + \mathbf{3} \times \mathbf{3})_{sym} + (\mathbf{3} \times \mathbf{1} + \mathbf{1} \times \mathbf{3})_{asym}$. $F$ is the anti-symmetric square of the vector which has two options $\mathbf{3} \times \mathbf{1}$ and $\mathbf{1} \times \mathbf{3}$ we call $F$ and $\bar{F}$ (that we know are actually the electric and magnetic fields). The Cartan charges for the $\mathbf{3}$ are $j = 0, \pm 1$. Note sometimes in the literature $F$ is written with vector representation indices $\pm\pm$ which is somewhat misleading [70][2].

- The charges of the vector $A_\mu$ are assigned to be the same as the derivative $\partial_\mu$, because they come together in a covariant derivative in a gauge theory. Therefore $\Delta(A_\mu) = \Delta(\partial_\mu) = 1$ in all dimensions. In the table above we refer to $F$ rather than $A$, which simply increases the dimension by one (since $F \sim \partial A$). Note that the kinetic term of the vector is of the form $\lambda F_{\mu\nu}^2$, and therefore the coupling constant $\lambda$ is dimensionless (similarly to the scalar field).

- For an interacting vector multiplet the R-charge assignment is simply $R(F) = R(A_\mu) = R(\partial_\mu) = 0$ (same as derivative), which is the same in all dimensions. The lowest component of the vector superfield is $A$, and the following component (with an extra $\theta$ factor in the superfield) is the anti-spinor $\bar{\lambda}$. The charges of the components are related with the operation of the supersymmetry generator $Q$ as $[\bar{\lambda}] = [\bar{Q}A]$. Therefore $R(\bar{\lambda}) = R(A) + R(\bar{Q}) = -1$ and $\Delta(\bar{\lambda}) = \Delta(A) + \Delta(\bar{Q}) = 1 + \frac{1}{2} = \frac{3}{2}$. The R-charges do not change in the RG flow because they are fixed by the R-charge of the derivative, so therefore $R_{IR} = R_{UV}$ for the vector components.

- Derivatives with $\delta = 0$ can be applied to any contributing operators to generate new contributing terms. The derivatives can be applied infinitely many times, giving the overall factor to an existing single-letter contribution $\sum_{k,l=0}^{\infty} \partial_{++}^{(k)} \partial_{-+}^{(l)} \to \sum_{k,l=0}^{\infty} p^k q^l = \frac{1}{(1-p)(1-q)}$.

- Note $\delta \geq 0$ for any operators, since $\langle 0|\{Q, Q\dagger\}|0\rangle = \langle 0|\delta|0\rangle \geq 0$. If it was not the case, we could combine $\delta < 0$ letters with $\delta > 0$ letters to give terms with $\delta = 0$ in total, which would also contribute to the index.

- The fields in the UV are free and therefore exactly satisfy their equations of motion (EOM). Since EOMs contain derivatives, this means that some of the derivatives (with $\delta = 0$), should actually not be counted in the index. Alternatively, we do count them, but also count the EOMs (with $\delta = 0$) with a negative sign. In case of the fermionic spinor field, the letter contributions will negative and the EOMs positive (double negative).

In the following table we write the charges and contributions of the EOMs, followed by explanations.

Table B.2: Charges and supersymmetric index contributions of EOMs in 4d $\mathcal{N}=1$

| Letter | Lorentz rep. | $\Delta$ | $j_1$ | $j_2$ | $R_{UV}$ | $R_{IR}$ | $\delta$ | $i$ |
|---|---|---|---|---|---|---|---|---|
| $\Box\phi = 0$ | $(0,0)$ | $\frac{8}{3}$ | $0$ | $0$ | $\frac{2}{3}$ | $r$ | $3$ | $0$ |
| $\bar{\sigma}^\mu \partial_\mu \bar{\psi}_+ = 0$ | $\left(\frac{1}{2},0\right)$ | $\frac{5}{2}$ | $\frac{1}{2}$ | $0$ | $\frac{1}{3}$ | $-r+1$ | $2$ | $0$ |
| $\sigma^\mu \partial_\mu \lambda_\pm = 0$ | $\left(0,\frac{1}{2}\right)$ | $\frac{5}{2}$ | $0$ | $\pm\frac{1}{2}$ | $1$ | $1$ | $0^+, 2^-$ | $-(-pq), 0$ |
| $\partial_\mu F^{\mu\nu} = 0$ | $\left(\frac{1}{2},\frac{1}{2}\right)$ | $3$ | $\pm\frac{1}{2}$ | $\pm\frac{1}{2}$ | $0$ | $0$ | $>0$ | $0$ |

Notes:

- The EOM of $\phi$ is the Klein-Gordon equation $\Box\phi = 0$. The object $\Box\phi$ is a scalar, and therefore has $j_1, j_2 = 0$. Its R-charge is the same as of $\phi$ (derivatives have zero R-charge) and its dimension is $\Delta = \Delta(\phi) + \Delta(\Box) = \frac{2}{3} + 2 = \frac{8}{3}$. Therefore, it has $\delta > 0$ and cannot contribute to the index.

- The EOM of a 4-component Dirac spinor is $\gamma^\mu \partial_\mu \psi = 0$, and for the 2-component spinors we have two Weyl equations $\sigma^\mu \partial_\mu \psi_\alpha = \bar{\sigma}^\mu \partial_\mu \bar{\psi}_\alpha = 0$, where $\sigma^0 = \bar{\sigma}^0 = \mathbf{1}_{2\times 2}$ and $\bar{\sigma}^i = -\sigma^i$. Define the derivative with spinor indices using the derivative in spacetime indices:

$$\partial_{\alpha\beta} = \sigma^\mu_{\alpha\beta}\partial_\mu = \sigma^0_{\alpha\beta}\partial_0 + \sigma^1_{\alpha\beta}\partial_1 + \sigma^2_{\alpha\beta}\partial_2 + \sigma^3_{\alpha\beta}\partial_3$$
$$\begin{pmatrix} \partial_{-+} & \partial_{++} \\ \partial_{--} & \partial_{+-} \end{pmatrix} = \begin{pmatrix} \partial_0 + \partial_3 & \partial_1 - i\partial_2 \\ \partial_1 + i\partial_2 & \partial_0 - \partial_3 \end{pmatrix} \tag{B.3}$$

We chose the labels $\alpha\beta$ to be consistent with the Lorentz charges, as we will immediately see. The other definition (relevant for the Weyl equation for $\bar{\psi}$) is:

$$\bar{\partial}_{\alpha\beta} = \bar{\sigma}^\mu_{\alpha\beta}\partial_\mu = \sigma^0_{\alpha\beta}\partial_0 - \left(\sigma^1_{\alpha\beta}\partial_1 + \sigma^2_{\alpha\beta}\partial_2 + \sigma^3_{\alpha\beta}\partial_3\right)$$
$$= \begin{pmatrix} \partial_0 - \partial_3 & -\partial_1 + i\partial_2 \\ -\partial_1 - i\partial_2 & \partial_0 + \partial_3 \end{pmatrix} = \begin{pmatrix} \partial_{+-} & -\partial_{++} \\ -\partial_{--} & \partial_{-+} \end{pmatrix} \tag{B.4}$$

The 4d vector representation in $SU(2) \times SU(2)$ is $\mathbf{2} \times \mathbf{2}$ with 4 components. Indeed we can from the definition with $\partial_\mu$ that all the spinor components are independent. The Weyl equation for $\psi$:

$$0 = \sigma^\mu \partial_\mu \psi = \begin{pmatrix} \partial_{-+} & \partial_{++} \\ \partial_{--} & \partial_{+-} \end{pmatrix} \begin{pmatrix} \psi_+ \\ \psi_- \end{pmatrix} = \begin{pmatrix} \partial_{-+}\psi_+ + \partial_{++}\psi_- \\ \partial_{--}\psi_+ + \partial_{+-}\psi_- \end{pmatrix} \tag{B.5}$$

The Weyl equation for $\bar{\psi}$:

$$0 = \bar{\sigma}^\mu \partial_\mu \bar{\psi} = \begin{pmatrix} \partial_{+-} & -\partial_{++} \\ -\partial_{--} & \partial_{-+} \end{pmatrix} \begin{pmatrix} \bar{\psi}_+ \\ \bar{\psi}_- \end{pmatrix} = \begin{pmatrix} \partial_{+-}\bar{\psi}_+ - \partial_{++}\bar{\psi}_- \\ -\partial_{--}\bar{\psi}_+ + \partial_{-+}\bar{\psi}_- \end{pmatrix} \quad (B.6)$$

The Lorentz charges of the spinors $\psi_\pm$ are $j_1(\psi_\pm) = \pm\frac{1}{2}, j_2(\psi_\pm) = 0$ (and the other way around for $\bar{\psi}$) we see that the assigning $j_1, j_2$ charges $\pm\frac{1}{2}$ for the derivatives consistent with the $\pm\pm$ signs will make the resulting EOM Lorentz charges also spinor as required.

As with the scalar, the R charges of the EOMs are the same as of the spinors, and the dimension is reduced by one $\Delta = \Delta(\psi) + \Delta(\partial) = \frac{5}{2}$. Checking the table above, we see that the EOMs for the spinors with $\delta = 0$ do not contribute to the index.

- The EOM of $F_{\mu\nu}$ is $\partial_\mu F^{\mu\nu} = 0$. The object $\partial_\mu F^{\mu\nu}$ is a vector, and therefore has $j_1 j_2 = \pm\frac{1}{2}$. Its R-charge is zero and its dimension is $\Delta = 3$. Therefore, it has $\delta > 0$ and cannot contribute to the index.

Overall, the single letter contribution of a chiral superfield:

$$i_\chi(q,p) = \frac{(pq)^{\frac{r}{2}} - (pq)^{\frac{2-r}{2}}}{(1-p)(1-q)} \quad (B.7)$$

Similarly, the single letter contribution of a vector superfield:

$$i_V(q,p) = \frac{2pq - p - q}{(1-p)(1-q)} \quad (B.8)$$

If the chiral superfield is also in a representation $R$ of some group $G$ (gauge group or a global symmetry group), then the field will also be charged with the fugacities $\mathbf{z}$ associated with the group $G$, which are the maximal torus variables (a set of rank $(G)$ parameters $\mathbf{z} = \{z_i\}_{i=1}^{\text{rank}(G)}$). For a field in a representation $R$ these fugacities will form the character $\chi_R(\mathbf{z})$ of that representation, given by $\sum_{\mu \in \text{wt}(R)} z^\mu$. See Appendix A.1 for more details. For example, the character of the fundamental of $U(1)$ is $\chi_R(\mathbf{z}) = z$ and of $SU(2)$ is $\chi_R(\mathbf{z}) = z + z^{-1}$.

The single-letter contribution of the field is then modified to:

$$i_\chi(q, p, \mathbf{z}) = \frac{(pq)^{\frac{r}{2}} \chi_R(\mathbf{z}) - (pq)^{\frac{2-r}{2}} \chi_{\bar{R}}(\mathbf{z})}{(1-p)(1-q)} \quad (B.9)$$

Note that $\chi_{\bar{R}}(\mathbf{z}) = \chi_R(\mathbf{z}^{-1})$. Same for the vector, that will necessarily be in the adjoint representation of a gauge group:

$$i_V(q, p, \mathbf{z}) = \frac{2pq - p - q}{(1-p)(1-q)} \chi_{\text{adj}}(\mathbf{z}) \quad (B.10)$$

Given we have the single-letter contributions to the index, the multi-letter (or full) contribution to the index is achieved by applying the plethystic exponential operation. First, let us define the plethystic exponential (PE) of some multi-variable function $g(\mathbf{t})$ (that vanishes at the origin $g(\mathbf{0}) = 0$) as

$$\text{PE}[g(\mathbf{t})] \equiv \exp\left(\sum_{n=1}^{\infty} \frac{g(\mathbf{t}^n)}{n}\right) \tag{B.11}$$

where by $\mathbf{t}^n$ we mean separately taking each of the variables of the function $g$ to the power $n$ [54][55][15][61]. Choosing $g$ to be the single-letter index, the PE performs the combinatorics involved with calculating the symmetrization of products of representations, meaning it is a generator for symmetrization [55][15]. Thus, applying the PE on the single letter gives the multi-letter index [18][15]:

$$I = \text{PE}[i(q, p, \mathbf{z})] = \exp\left\{\sum_{m=1}^{\infty} \frac{1}{m} i(q^m, p^m, \mathbf{z}^m)\right\} \tag{B.12}$$

If the group $G$ is a gauge group (rather than a global symmetry group) the maximal torus variables $\mathbf{z}$ of the group describe redundancies in the theory. Only gauge invariant states (singlets of the gauge group) can contribute to the index. Therefore, we need to project the index on the trivial representation of $G$. The projection is done by integrating the maximal torus variables over the group $G$ with the Haar measure of the group [55][42]:

$$I(q, p) = \int_G d\mu_G(\mathbf{z}) \, \text{PE}[i(q, p, \mathbf{z})] \tag{B.13}$$

For more details on the Haar measure of different Lie groups see Appendix A.2.

For $U(1)$ global symmetry with fugacity $z$, the muti-letter contribution of the chiral

multiplet is

$$
\begin{aligned}
I_\chi(p,q,z) &= \exp\left\{\sum_{m=1}^\infty \frac{1}{m} i_\chi(q^m, p^m, z^m)\right\} \\
&= \exp\left\{\sum_{m=1}^\infty \frac{1}{m} \frac{\left((pq)^{\frac{r}{2}} z\right)^m - \left((pq)^{\frac{2-r}{2}} z^{-1}\right)^m}{(1-p^m)(1-q^m)}\right\} \\
&= \exp\left\{\sum_{m=1}^\infty \frac{1}{m}\left[\left((pq)^{\frac{r}{2}} z\right)^m - \left((pq)^{\frac{2-r}{2}} z^{-1}\right)^m\right] \sum_{k,l=0}^\infty p^{mk} q^{ml}\right\} \\
&= \exp\left\{\sum_{k,l=0}^\infty \sum_{m=1}^\infty \frac{1}{m}\left[\left((pq)^{\frac{r}{2}} z p^k q^l\right)^m - \left((pq)^{\frac{2-r}{2}} z^{-1} p^k q^l\right)^m\right]\right\} \quad \text{(B.14)} \\
&= \exp\left\{-\sum_{k,l=0}^\infty \left[\ln\left(1 - (pq)^{\frac{r}{2}} z p^k q^l\right) - \ln\left(1 - (pq)^{\frac{2-r}{2}} z^{-1} p^k q^l\right)\right]\right\} \\
&= \prod_{k,l=0}^\infty \exp\left\{-\ln\left(1 - (pq)^{\frac{r}{2}} z p^k q^l\right) + \ln\left(1 - (pq)^{\frac{2-r}{2}} z^{-1} p^k q^l\right)\right\} \\
&= \prod_{k,l=0}^\infty \frac{1 - (pq)^{\frac{2-r}{2}} z^{-1} p^k q^l}{1 - (pq)^{\frac{r}{2}} z p^k q^l} = \prod_{k,l=0}^\infty \frac{1 - (pq)^{-\frac{r}{2}} z^{-1} p^{k+1} q^{l+1}}{1 - (pq)^{\frac{r}{2}} z p^k q^l} \\
&= \Gamma\left((pq)^{\frac{r}{2}} z; p, q\right)
\end{aligned}
$$

where we used the Taylor expansion $\sum_{m=1}^\infty \frac{z^m}{m} = -\ln(1-z)$, and the definition of the elliptic Gamma function:

$$\Gamma(z; p, q) \equiv \prod_{k,l=0}^\infty \frac{1 - z^{-1} p^{k+1} q^{l+1}}{1 - z p^k q^l} \quad \text{(B.15)}$$

For a general representaion, the multi-letter contribution of the chiral will be gen-

eralized to:

$$I_\chi = \exp\left\{\sum_{m=1}^{\infty} \frac{1}{m} i_\chi(q^m, p^m, \mathbf{z}^m)\right\}$$

$$= \exp\left\{\sum_{m=1}^{\infty} \frac{1}{m} \frac{\left((pq)^{\frac{r}{2}}\right)^m \chi_R(\mathbf{z}^m) - \left((pq)^{\frac{2-r}{2}}\right)^m \chi_R(\mathbf{z}^{-m})}{(1-p^m)(1-q^m)}\right\}$$

$$= \exp\left\{\sum_{m=1}^{\infty} \frac{1}{m} \frac{\left((pq)^{\frac{r}{2}}\right)^m \sum_{\mu \in \mathrm{wt}(R)} (z^\mu)^m - \left((pq)^{\frac{2-r}{2}}\right)^m \sum_{\mu \in \mathrm{wt}(R)} (z^\mu)^{-m}}{(1-p^m)(1-q^m)}\right\} \quad \text{(B.16)}$$

$$= \exp\left\{\sum_{\mu \in \mathrm{wt}(R)} \sum_{m=1}^{\infty} \frac{1}{m} \frac{\left((pq)^{\frac{r}{2}} z^\mu\right)^m - \left((pq)^{\frac{2-r}{2}} z^{-\mu}\right)^m}{(1-p^m)(1-q^m)}\right\}$$

$$= \prod_{\mu \in \mathrm{wt}(R)} I_\chi(p, q, z^\mu)$$

Notice how the character which is a sum $\sum_{\mu \in \mathrm{wt}(R)}$ transforms to a product $\prod_{\mu \in \mathrm{wt}(R)}$ in the level of the multi-letter index.

In the same manner, the multi-letter contribution of the vector (skipping some of the steps that were the same for the chiral superfield) is

$$I_V = \exp\left\{\sum_{m=1}^{\infty} \frac{1}{m} i_V(q^m, p^m, \mathbf{z}^m)\right\}$$

$$= \exp\left\{\sum_{m=1}^{\infty} \frac{1}{m} \frac{2(pq)^m - p^m - q^m}{(1-p^m)(1-q^m)} \chi_{adj}(\mathbf{z}^m)\right\}$$

$$= \prod_{\mu \in \mathrm{wt}(R)} \exp\left\{\sum_{m=1}^{\infty} \frac{1}{m} \frac{2(pq)^m - p^m - q^m}{(1-p^m)(1-q^m)} (z^\mu)^m\right\} \quad \text{(B.17)}$$

$$= \prod_{\mu \in \mathrm{wt}(R)} \exp\left\{-\sum_{k,l=0}^{\infty} [2\ln(1-pqy) - \ln(1-py) - \ln(1-qy)]\right\}$$

$$= \prod_{\mu \in \mathrm{wt}(R)} \prod_{k,l=0}^{\infty} \frac{(1-py)(1-qy)}{(1-pqy)^2}$$

where $y \equiv p^k q^l z^\mu$.

## B.3   3d $\mathcal{N} = 2$ index

Here we will calculate the 3d index in the zero monopole sector. The contribution of monopoles is not trivial and has to be calculated using localization. The vector contribution cannot be calculated in any case using the trace formula, as we explain below.

The 3d $\mathcal{N} = 2$ multiplets are [25]:

- $\mathcal{N} = 2$ chiral multiplet, containing a scalar field $\phi$ and a spinor field $\psi_\pm$ (both complex).

- $\mathcal{N} = 2$ vector multiplet, containing the vector field $A_\mu$ (or the gauge-invariant field strength tensor $F_{\mu\nu}$) and a spinor field $\lambda_\pm$, similar to the analogue 4d $\mathcal{N} = 1$ vector multiplet. A difference is the additional real scalar $\sigma$ (in the same place in the superfield as $A_\mu$) which can be thought of as the fourth component of $A_\mu$ in the 4d case.

- It is important to note here that the kinetic term of the vector is of the form $\lambda F_{\mu\nu}^2$, and therefore the coupling constant $\lambda$ is dimensionful in 3d. This means that the vector has a scale as it flows to the UV, which forbids the existance of a superconformal vector multiplet in 3d. Since we are calculating the index in superconformal fixed points, it is not possible to define the vector contribution to the index as a trace formula (counting letters) in 3d.

The 3d $\mathcal{N} = 2$ index is [16]

$$I = \text{Tr}_\mathcal{H} \left( (-1)^F e^\delta q^{j_3 + \frac{R}{2}} \mu^F \right)$$
$$\delta = \Delta - j_3 - R$$
(B.18)

where $\Delta$ the dilatation generator, $j_3$ the $SO(3) \sim SU(2)$ Lorentz group Cartan subalgebra generator and $R$ the $U(1)$ R-symmetry generator.

Similarly to the previous section for 4d $\mathcal{N} = 1$, in the following table we write down the available letters, their charges with respect to the different generators, and their contributions to the (single letter) index $i$ in case they satisfy $\delta = 0$, according to Eq. B.18. In the notes that follow we provide detailed explanations.

Table B.3: Charges and supersymmetric index contributions of different letters in 3d $\mathcal{N}=2$

| Letter | $\Delta$ | $j_3$ | $R_{UV}$ | $R_{IR}$ | $\delta$ | $i$ |
|---|---|---|---|---|---|---|
| $Q_\pm$ | $\frac{1}{2}$ | $\pm\frac{1}{2}$ | 1 | | | |
| $\phi$ | $\frac{1}{2}$ | 0 | $\frac{1}{2}$ | $r$ | 0 | $q^{\frac{r}{2}}$ |
| $\bar{\phi}$ | $\frac{1}{2}$ | 0 | $-\frac{1}{2}$ | $-r$ | 1 | 0 |
| $\psi_\pm$ | 1 | $\pm\frac{1}{2}$ | $-\frac{1}{2}$ | $r-1$ | $1^+, 2^-$ | 0 |
| $\bar{\psi}_\pm$ | 1 | $\pm\frac{1}{2}$ | $\frac{1}{2}$ | $-r+1$ | $0^+, 1^-$ | $-q^{\frac{2-r}{2}}, 0$ |
| $\partial_\pm$ | 1 | $\pm 1$ | 0 | 0 | $0^+, 2^-$ | $q, 0$ |
| $\partial_0$ | 1 | 0 | 0 | 0 | 1 | 0 |

Notes:

- The supersymmetry generator $Q_\pm$ has the same charges as in 4d. Also, in 3d there is just one spinor, unlike the right-handed or left-handed Weyl spinors in 4d.

- As in 4d, we choose the dimensions of the chiral field components using their canonical values $[\phi] = \frac{d-2}{2} = \frac{1}{2}, [\psi] = \frac{d-1}{2} = 1$. The $R_{UV}$ charge of $\phi$ is assigned using the unitarity bound $\Delta = |R_{UV}|$ in 3d (different bound than in 4d), and the $R_{UV}$ charge of $\psi$ in the same manner as in the previous section giving $R_{UV}(\psi) = R_{UV}(\phi) - 1$.

- The derivative with $\delta = 0$ is $\partial_+$. As explained in the previous section, it can be applied infinitely to give an overall factor of $\sum_{k=0}^\infty \partial_+^{(k)} \to \sum_{k=0}^\infty q^l = \frac{1}{(1-q)}$.

In the following table we write the charges and contributions of the EOMs, followed by explanations.

Table B.4: Charges and supersymmetric index contributions of EOMs in 3d $\mathcal{N}=2$

| Letter | $\Delta$ | $j_3$ | $R_{UV}$ | $R_{IR}$ | $\delta$ | $i$ |
|---|---|---|---|---|---|---|
| $\Box\phi = 0$ | $\frac{5}{2}$ | 0 | $\frac{1}{2}$ | $r$ | 2 | 0 |
| $\partial_{-+}\bar{\psi}_+$ | 2 | $\frac{1}{2}$ | $\frac{1}{2}$ | $-r+1$ | 1 | 0 |
| $\partial_{--}\bar{\psi}_+$ | 2 | $-\frac{1}{2}$ | $\frac{1}{2}$ | $-r+1$ | 2 | 0 |

Notes:

- Same as the previous section, The EOM of $\phi$ is $\Box\phi = 0$. The object $\Box\phi$ is a scalar $j_3 = 0$, the R-charge is $R_{UV}(\Box\phi) = R_{UV}(\phi) = \frac{1}{2}$ and the dimension $\Delta(\Box\phi) = \Delta(\phi) + \Delta(\Box) = \frac{1}{2} + 2 = \frac{5}{2}$. In total $\delta > 0$ and it does not contribute to the index.

- The EOM of a spinor $\psi$ (which has two components) in 3d is the Weyl equation $\sigma^\mu \partial_\mu \psi_\alpha = 0$, where $\mu = 1, 2, 3$ are the spacetime indices and $\sigma_i$ the three Pauli matrices (the identity matrix is not required in 3d to satisfy the Clifford algebra). Define the derivative with spinor indices using the derivative in spacetime indices:

$$\partial_{\alpha\beta} = \sigma^\mu_{\alpha\beta} \partial_\mu = \sigma^1_{\alpha\beta} \partial_1 + \sigma^2_{\alpha\beta} \partial_2 + \sigma^3_{\alpha\beta} \partial_3$$
$$\begin{pmatrix} \partial_{-+} & \partial_{++} \\ \partial_{--} & \partial_{+-} \end{pmatrix} = \begin{pmatrix} \partial_3 & \partial_1 - i\partial_2 \\ \partial_1 + i\partial_2 & -\partial_3 \end{pmatrix} \tag{B.19}$$

We chose the labels $\alpha\beta$ to be consistent with the Lorentz charges, as we will immediately see. The 3d vector representation of $SU(2)$ is **3**, which can also be written as $(\mathbf{2} \times \mathbf{2})_{sym}$. Therefore, only 3 of the components in the matrix above are independent. We can see with the definition using the derivative in spacetime indices $\partial_\mu$ that $\partial_{-+} = -\partial_{+-}$, and the other two components are independent (they are complex conjugates).

The Weyl equation for $\psi$ (same as previous section):

$$0 = \sigma^\mu \partial_\mu \psi = \begin{pmatrix} \partial_{-+} & \partial_{++} \\ \partial_{--} & \partial_{+-} \end{pmatrix} \begin{pmatrix} \psi_+ \\ \psi_- \end{pmatrix} = \begin{pmatrix} \partial_{-+}\psi_+ + \partial_{++}\psi_- \\ \partial_{--}\psi_+ + \partial_{+-}\psi_- \end{pmatrix} \tag{B.20}$$

The Lorentz charge $j_3$ of the spinor components $\psi_\pm$ are $j_3(\psi_\pm) = \pm\frac{1}{2}$, and therefore the correct assignment of the charges of the derivative components such that the Weyl equation remains a spinor is $j_3(\partial_{++}) = -j_3(\partial_{--}) = 1$ and $j_3(\partial_{-+}) = j_3(\partial_{+-}) = 0$. The correct mapping between the $(\mathbf{2} \times \mathbf{2})_{sym}$ notation and the **3** notation is therefore:

$$\partial_{++} \leftrightarrow \partial_+$$
$$\partial_{--} \leftrightarrow \partial_- \tag{B.21}$$
$$\partial_{-+}, \partial_{+-} \leftrightarrow \partial_0$$

As with the scalar, the R charges of the EOMs are the same as of the spinors, and the dimension is reduced by one. In the table above we write the EOMs of the letters that had $\delta = 0$ and we find that all the EOMs have $\delta > 0$ and therefore do not contribute to the index.

- In total, no EOM contributions in 3d.

The chiral superfield single-letter contribution is:

$$i_\chi(q) = \frac{q^{\frac{r}{2}} - q^{\frac{2-r}{2}}}{(1-q)} \tag{B.22}$$

For a general representation:

$$i_\chi(q, \mathbf{z}) = \frac{q^{\frac{r}{2}} \chi_R(\mathbf{z}) - q^{\frac{2-r}{2}} \chi_{\bar{R}}(\mathbf{z})}{(1-q)} \tag{B.23}$$

The multi-letter contribution of the chiral multiplet (again, for a fundamental of $U(1)$ representation) is

$$\begin{aligned}
I_\chi(q, z) &= \exp\left\{ \sum_{m=1}^\infty \frac{1}{m} i_\chi(q^m, z^m) \right\} \\
&= \exp\left\{ \sum_{m=1}^\infty \frac{1}{m} \frac{\left(q^{\frac{r}{2}} z\right)^m - \left(q^{\frac{2-r}{2}} z^{-1}\right)^m}{(1-q^m)} \right\} \\
&= \exp\left\{ \sum_{m=1}^\infty \frac{1}{m} \left[\left(q^{\frac{r}{2}} z\right)^m - \left(q^{\frac{2-r}{2}} z^{-1}\right)^m\right] \sum_{l=0}^\infty q^{ml} \right\} \\
&= \exp\left\{ \sum_{l=0}^\infty \sum_{m=1}^\infty \frac{1}{m} \left[\left(q^{\frac{r}{2}} z p^k q^l\right)^m - \left(q^{\frac{2-r}{2}} z^{-1} p^k q^l\right)^m\right] \right\} \\
&= \exp\left\{ -\sum_{k,l=0}^\infty \left[\ln\left(1 - q^{\frac{r}{2}} z p^k q^l\right) - \ln\left(1 - q^{\frac{2-r}{2}} z^{-1} p^k q^l\right)\right] \right\} \\
&= \prod_{l=0}^\infty \frac{1 - q^{\frac{2-r}{2}} z^{-1} q^l}{1 - q^{\frac{r}{2}} z q^l} = \frac{\left(q^{\frac{2-r}{2}} z^{-1}; q\right)}{\left(q^{\frac{r}{2}} z; q\right)}
\end{aligned} \tag{B.24}$$

where the definition of the $q$-Pochhammer symbol is:

$$(z; q) \equiv \prod_{l=0}^\infty \left(1 - zq^l\right) \tag{B.25}$$

As in the 4d case of the last section, for a general representation $R$ the multi-letter index of the chiral multiplet is generalized to:

$$I_\chi(q, \mathbf{z}) = \prod_{\mu \in \text{wt}(R)} I_\chi(q, z^\mu) \tag{B.26}$$

### B.3.1 Treating non-zero monopole sectors

We also note that in a non-zero monopole sector, each maximal torus variable $z_i$ (fugacity) has a corresponding magnetic monopole charge $m_i$ (flux) [11][15]. For a general representation $R$, the maximal torus variable $z$ is generalized to $z^\mu$ (shorthand notation for $\prod_{j=1}^{\text{rank}(G)} z_j^{\mu_{i,j}}$), where $\mu = \left(\mu_1, ..., \mu_{\text{rank}(G)}\right)$ is a weight vector of the representation (see Appendix A.1). The magnetic monopole charge $m$ (corresponding to $z$) is generalized to $\mu(m)$ (shorthand notation for the scalar product $\vec{\mu} \cdot \vec{m}$, where $\vec{m} = \left(m_1, ..., m_{\text{rank}(G)}\right)$) [16].

Therefore, given we know the basic building block $I_\chi(q, z, m)$ (see chapter 3.2), the generalization to the multi-letter index is [11]:

$$I_\chi(q, \mathbf{z}, \mathbf{m}) = \prod_{\mu \in \text{wt}(R)} I_\chi(q, z^\mu, \mu(m)) \tag{B.27}$$

For the Lie group $SU(N)$ with rank $N-1$ it is still convenient sometimes to treat its weight vector as $N$ dimensional with the constraint $\prod_{i=1}^N z_i = 1$. Same for the magnetic monopole charges with the constraint $\sum_{i=1}^N m_i = 0$ (consistent with the sum of powers of the $z_i$'s equal to zero).

The contribution of the vector multiplet (in the zero or non-zero monopole sectors) is calculated using localization (not derived in this thesis) and the result is written in chapter 3.2 (Eq. 3.6). The result depends on the root vectors $\alpha$ (see Appendix A.2) and the notation $\alpha(m)$ is shorthand for $\vec{\alpha} \cdot \vec{m}$, similar to the notation $\mu(m)$.

## B.4   3d $\mathcal{N} = 4$ index

Moving on to $\mathcal{N} = 4$ supersymmetry, we have an enlarged R-symmetry $G_R \equiv SU(2)_H \times SU(2)_C$ with the two corresponding Cartan subalgebra generators $R_H, R_C$. Below, we will use the index $M$ for $SU(2)_H$, the index $A$ for $SU(2)_C$ and the index $\alpha$ for the Lorentz group $SU(2)$.

The 3d $\mathcal{N} = 4$ multiplets are [11]:

- $\mathcal{N} = 4$ hypermultiplet, containing two $\mathcal{N} = 2$ chiral multiplets $\Phi_1, \Phi_2$ with conjugate global/flavor and gauge symmetry representations ($\Phi_1, \Phi_2^\dagger$ have the same flavor and gauge representations). The (complex) scalar fields of both chiral multiplets $\phi_1, \phi_2^\dagger$ form a doublet representation of $SU(2)_H$ (($\mathbf{2}, \mathbf{1}$) representation of $G_R$) that we call $\phi_M$, and the spinor fields $\psi_1, \psi_2^\dagger$ form a doublet representation of $SU(2)_C$ (($\mathbf{1}, \mathbf{2}$) representation of $G_R$) that we call $\psi_{\alpha, A}$. The complex conjugates of $\phi_M, \psi_{\alpha, A}$ are in the same representations of $G_R$ (since the representations of $SU(2)$ are all real) but in the conjugate representations of the global symmetry group.

- $\mathcal{N} = 4$ vector multiplet, containing an $\mathcal{N} = 2$ chiral multiplet $\Phi$ and an $\mathcal{N} = 2$

vector multiplet $V_\mu$ in the adjoint representation of the gauge group. The spinor fields of the chiral multiplet $\lambda_1$ and the vector multiplet $\lambda_2$ form together the $(\mathbf{2}, \mathbf{2})$ representation of $G_R$, the vector $A_\mu$ of the vector multiplet is a singlet of $G_R$ and the real scalar field $\sigma$ of the vector multiplet together with the complex scalar field of the chiral multiplet $\varphi$ (three real scalar components in total) form a triplet of $SU(2)_C$ ($(\mathbf{1}, \mathbf{3})$ representation of $G_R$) that we call $\sigma_A$.

- As noted in the previous section, we cannot use the trace formula for the vector. The contribution is calculated using localization and the result appears in chapter 3.2.

For $\mathcal{N} = 4$ the index is [11][16]

$$I = \text{Tr}_\mathcal{H} \left( (-1)^F e^\delta q^{j_3 + \frac{1}{2}(R_H + R_C)} t^{R_H - R_C} \mu^F \right)$$
$$\delta = \Delta - j_3 - R_H - R_C$$
(B.28)

where $\Delta$ the dilatation generator, $j_3$ the $SO(3) \sim SU(2)$ Lorentz group Cartan subalgebra generator and $R_H, R_C$ the $SU(2)_H \times SU(2)_C$ R-symmetry Cartan subalgebra generators. The dependency of $\delta$ in $R_H, R_C$ dictates how the other operators look like, due to the constraints of commutation with $\delta$ and independency. We can see the dependency is in the combinations $R_H \pm R_C$ only, which means the R-symmetry group is described in terms of the $U(1) \times U(1)$ subgroup generators (see Appendix A.4).

Similarly to the previous sections, in the following table we write down the letters and their contributions, according to Eq. B.28.

Table B.5: Charges and supersymmetric index contributions of different letters in 3d $\mathcal{N} = 4$

| Letter | R-sym. rep. | $\Delta$ | $j_3$ | $R_{UV}$ | $R_{IR}$ | $\delta$ | $i$ |
|---|---|---|---|---|---|---|---|
| $Q$ | $\mathbf{2} \times \mathbf{2}$ | $\frac{1}{2}$ | $\pm\frac{1}{2}$ | $\pm\frac{1}{2}$ | $\pm\frac{1}{2}$ | | |
| $\phi_M$ | $\mathbf{2} \times \mathbf{1}$ | $\frac{1}{2}$ | $0$ | $\pm\frac{1}{2}, 0$ | $0$ | $0^+, 1^-$ | $q^{\frac{1}{4}} t^{\frac{1}{2}}, 0$ |
| $\psi_{+,A}$ | $\mathbf{1} \times \mathbf{2}$ | $1$ | $\frac{1}{2}$ | $0$ | $\pm\frac{1}{2}, 0$ | $0^+, 1^-$ | $q^{\frac{3}{4}} t^{-\frac{1}{2}}, 0$ |
| $\partial_\pm$ | $\mathbf{1} \times \mathbf{1}$ | $1$ | $\pm 1$ | $0$ | $0$ | $0^+, 2^-$ | $q, 0$ |
| $\partial_0$ | $\mathbf{1} \times \mathbf{1}$ | $1$ | $0$ | $0$ | $0$ | $1$ | $0$ |

Notes:

- The notes from the previous section still apply, except the R-charges are assigned using the fact we know the $G_R$ representations of the letters.

- The EOMs are also the same as the previous section, and they do not contribute.

- The R-charges do not change in the flow from the UV to the IR because no-mixing is allowed for non-abelian R-symmetry (also tested with Z-extremization for some examples in section 3.5).

The single letter contribution of the hypermultiplet has a contribution from $\phi_+, \psi_{+,+}$ and also their conjugates $\bar{\phi}_+, \bar{\psi}_{+,+}$ ($z$ is a global symmetry fugacity):

$$i_{\text{hyp}}(q,t,z) = \frac{q^{\frac{1}{4}}t^{\frac{1}{2}}(z+z^{-1}) - q^{\frac{3}{4}}t^{-\frac{1}{2}}(z+z^{-1})}{(1-q)} \tag{B.29}$$

Therefore, in the multi-letter contribution it will give a contribution of two $\mathcal{N}=2$ chirals with opposite global symmetry charge, but with modified $q,t$ powers:

$$\begin{aligned}
I_{\text{hyp}}(q,t,z) &= \exp\left\{\sum_{m=1}^{\infty} \frac{1}{m} i_{\text{hyp}}(q^m, t^m, z^m)\right\} \\
&= \exp\left\{\sum_{m=1}^{\infty} \frac{1}{m} \frac{\left(q^{\frac{1}{4}}t^{\frac{1}{2}}(z+z^{-1})\right)^m - \left(q^{\frac{3}{4}}t^{-\frac{1}{2}}(z+z^{-1})\right)^m}{(1-q^m)}\right\} \\
&= \prod_{l=0}^{\infty} \frac{1-q^{\frac{3}{4}}t^{-\frac{1}{2}}zq^l}{1-q^{\frac{1}{4}}t^{\frac{1}{2}}zq^l} \frac{1-q^{\frac{3}{4}}t^{-\frac{1}{2}}z^{-1}q^l}{1-q^{\frac{1}{4}}t^{\frac{1}{2}}z^{-1}q^l} \\
&= \frac{\left(q^{\frac{3}{4}}t^{-\frac{1}{2}}z;q\right)\left(q^{\frac{3}{4}}t^{-\frac{1}{2}}z^{-1};q\right)}{\left(q^{\frac{1}{4}}t^{\frac{1}{2}}z;q\right)\left(q^{\frac{1}{4}}t^{\frac{1}{2}}z^{-1};q\right)} = \frac{\left(q^{\frac{3}{4}}t^{-\frac{1}{2}}z^{\pm 1};q\right)}{\left(q^{\frac{1}{4}}t^{\frac{1}{2}}z^{\pm 1};q\right)}
\end{aligned} \tag{B.30}$$

This result is consistent with the corresponding formula appearing in [11].

# Appendix C

# Conformal Manifold Supplementary

## C.1 Effect of an enhanced symmetry on the DCM

A remaining question is what happens to the DCM if we misidentify the global symmetry in the IR and it is actually enhanced from $G$ to a larger $\tilde{G}$ (meaning $G \subset \tilde{G}$). In this case we are not using the correct fugacities to describe the symmetry.

Generally we cannot say what the effect will be. In the special case the enhanced symmetry is a product group $\tilde{G} = G \times G'$ such that $G'$ is non-abelian, the extra conserved current might hide in the $q$ coefficient of the supersymmetric index as following:

$$I_q = \chi_{\text{adj}(G')} + \sum_i \chi_{R_i(G)}^{\text{marg}} - \chi_{\text{adj}(G)} - \chi_{\text{adj}(G')} \tag{C.1}$$

Meaning, the extra conserved current is compensated by a marginal operator in the adjoint representation. Using the large representation formula for the DCM from section 4.4, the DCM will not be affected in this case. This conclusion might break down in the original representations of $G$ are small, see examples in Appendix C.3.

If $G' = U(1)$ is abelian, then the index might take the more complicated form

$$I_q = a^Q + a^{Q'} \sum_i \chi_{R_i(G)}^{\text{marg}} - \chi_{\text{adj}(G)} - 1 \tag{C.2}$$

where $a$ is the fugacity of the additional $U(1)$ symmetry, and $Q, Q'$ are the $U(1)$ charges of the different operators (can be generalized such that each representation $R_i$ is charged with a different $Q_i'$). Being unaware of this extra symmetry is equivalent to taking the fugacities to one, and indeed plugging $a = 1$ restores the original index.

The DCM in this case will depend on the values of $Q, Q'$:

- $Q = 0, Q' \neq 0$ or $Q \neq 0, Q' = 0$: no way to create singlets of the symmetry by multipying the marginal operator representations, and therefore DCM = 0.

- $Q \neq 0, Q' \neq 0$ and $Q \cdot Q' > 0$: same as last case, no way to create singlets and therefore DCM = 0.

- $Q \neq 0, Q' \neq 0$ and $Q \cdot Q' < 0$: multiplying a representation $a^{Q'} \chi_{R_i(G)}^{marg}$ with $a^Q$ can form a singlet of $G'$ and therefore the DCM is no longer zero. For large representations the DCM conforms with the formula from section 4.4 and thus is not affected. Again, this can break down for small representations, see examples in Appendix C.3.

## C.2 Hilbert series implementation in Mathematica

In section 4.3 we defined an algorithm to calculate the Hilbert series in the general case of several representations of some Lie group. Here, we would like to describe how we implemented the algorithm in Mathematica [52] in its various stages:

1. Decomposition of the denominator to monomials is done using the List@@ command.

2. Calculation of zeros of each monomial is done using the Solve command.

3. Deletion of duplicates in the set of pole candidates is done using the DeleteDuplicates command.

4. Calculation of residues is done using the Residue command.

5. The progression of the algorithm for the $z_{i+1}$ variable can be continued in two alternative ways:

    (a) "Simplify" method: sum all the resulting residues $\sum_i \text{res}_i$ and try to simplify the expression (FullSimplify command). Then repeat the algorithm for integration over the $z_{i+1}$ variable. The problem here is that simplification of expressions can be prohibitive in complicated scenarios.

    (b) "Branches" method: repeat the algorithm for integration over the $z_{i+1}$ variable for each resulting residue expression $\text{res}_i$ separately (the name of the method refers to the branching/tree structure of the calculation). Once the final integration over $z_i$ for $i = \text{rank}(G)$ in done, sum all the resulting expressions. An additional improvement was added to this algorithm: in case some of the residue calculations fail (see the notes in the following paragraph) we can still try to save the calculation by going back to a level where the residues were fine, sum and simplify only the terms whose "offsprings" were faulty, and then repeat the procedure. This can be thought of as partially "downgrading" the Branches method to the Simplify method.

Notes:

- The number of residue calculations necessary at every level of the calculation (every integration variable) is given by the number of poles, which is roughly the number of monomials in the denominator, or the dimension of the representation $\dim(R)$. As the level of calculation progresses this process is repeated $\text{rank}(G)$ times. Assuming that the number of residue calculations remains roughly the same in each level, the complexity of the calculation is $O\left(\dim(R)^{\text{rank}(G)}\right)$. However, this naive complexity assesment does not take into account the fact that calculating the residue or simplifying expressions is (possibly exponentialy) harder when more symbolic variables are involved (being the hardest in the beginning of the calculation).

- Each residue calculation (in both methods) and each seperate branch in the Branches method can in principle be parallelized to speed up the calculations, but it was not implemented in this thesis.

- In the Branches method, there is a possibility that some of the residue calculations might fail or give a non-finite result, which would invalidate the entire calculation. This does not mean the original expression is problematic, because if we were to sum the problematic terms prior to the evaluation of the residues it would not have happened. In such cases we have to resort to using the Simplify method which is theoretically immune to this problem (although prohibitive in complicated cases).

- Even in the Simplify method a residue calculation might fail simply because Mathematica cannot handle it. This will be the main obstacle we face when we try to calculate the dimension in the general case.

- In a successful Branches method calculation (or the in last level of a Simplify method calculation), the result is a (possibly large) set of terms $\{\mathcal{H}_i(r)\}$ where the full Hilbert series is the sum of all the terms $\mathcal{H}(r) = \sum_i \mathcal{H}_i(r)$. As was already mentioned, summing and simplifying many terms can be prohibitive, but it is required only if we are interested in the exact analytic form of the Hilbert series. The dimension of the manifold can be deduced by numerically evaluting the sum for different values of $r$ approaching 1 (the leading pole will dominate in the limit $r \to 1$). The dimension will be the slope of the linear fit of $\ln(\mathcal{H}(r))$ as a function of $r$.

The exact implementation can be found in the accompanying Mathematica package [53].

## C.3  Manifold dimensions for different Lie groups

In chapter 4.3 we described an algorithm that allows to calculate the dimension of the manifold spanned by a set of operators given in representations of some Lie group. Here, we apply the algorithm for several Lie groups: $SU(2)$, $SU(3)$, $SU(4)$, $SO(7)$, $USp(4)$, $USp(6)$ and also their product with $U(1)$ (noted by the maximal torus variable $t$). For each group we perform the calculation for the smallest representations and work our way up to more complex cases.

Table C.1: Dimensions of the manifolds spanned by different representations of $SU(2)$ and $SU(2) \times U(1)$

| Group | Representation | Dimension |
|---|---|---|
| $SU(2)$ | $\chi_{\mathbf{2}}^{SU(2)}$ | 0 |
| $SU(2)$ | $\chi_{\mathbf{2}}^{SU(2)} + \chi_{\mathbf{2}}^{SU(2)}$ | 1 |
| $SU(2)$ | $\chi_{\mathbf{3}}^{SU(2)}$ | 1 |
| $SU(2)$ | $\chi_{\mathbf{3}}^{SU(2)} + \chi_{\mathbf{3}}^{SU(2)}$ | 3 |
| $SU(2)$ | $\chi_{\mathbf{4}}^{SU(2)}$ | 1 |
| $SU(2)$ | $\chi_{\mathbf{5}}^{SU(2)}$ | 2 |
| $SU(2)$ | $\chi_{\mathbf{6}}^{SU(2)}$ | 3 |
| $SU(2)$ | $\chi_{\mathbf{7}}^{SU(2)}$ | 4 |
| $SU(2) \times U(1)$ | $\chi_{\mathbf{2}}^{SU(2)} t + \frac{1}{t}$ | 0 |
| $SU(2) \times U(1)$ | $\chi_{\mathbf{2}}^{SU(2)} t + \chi_{\mathbf{2}}^{SU(2)} \frac{1}{t}$ | 1 |
| $SU(2) \times U(1)$ | $\chi_{\mathbf{3}}^{SU(2)} t + \frac{1}{t}$ | 1 |
| $SU(2) \times U(1)$ | $\chi_{\mathbf{4}}^{SU(2)} t + \frac{1}{t}$ | 1 |

To clarify, $\chi_{\mathbf{2}}^{SU(2)} t + \frac{1}{t}$ means we have a combination of the fundamental $\mathbf{2}$ and singlet $\mathbf{1}$ representations of $SU(2)$, where each representation is charged $\pm 1$ with $U(1)$.

Table C.2: Dimensions of the manifolds spanned by different representations of $SU(3)$ and $SU(3) \times U(1)$

| Group | Representation | Dimension |
|---|---|---|
| $SU(3)$ | $\chi_{\mathbf{3}}^{SU(3)}$ | 0 |
| $SU(3)$ | $2 \cdot \chi_{\mathbf{3}}^{SU(3)} = \chi_{\mathbf{3}}^{SU(3)} + \chi_{\mathbf{3}}^{SU(3)}$ | 0 |
| $SU(3)$ | $3 \cdot \chi_{\mathbf{3}}^{SU(3)}$ | 1 |
| $SU(3)$ | $4 \cdot \chi_{\mathbf{3}}^{SU(3)}$ | 4 |
| $SU(3)$ | $5 \cdot \chi_{\mathbf{3}}^{SU(3)}$ | 7 |
| $SU(3)$ | $\chi_{\mathbf{3}}^{SU(3)} + \chi_{\mathbf{\bar{3}}}^{SU(3)}$ | 1 |
| $SU(3)$ | $\chi_{\mathbf{6}}^{SU(3)}$ | 1 |
| $SU(3)$ | $\chi_{\mathbf{6}}^{SU(3)} + \chi_{\mathbf{6}}^{SU(3)}$ | 4 |
| $SU(3)$ | $\chi_{\mathbf{8}}^{SU(3)}$ | 4 |
| $SU(3)$ | $\chi_{\mathbf{8}}^{SU(3)}$ | 2 |
| $SU(3)$ | $\chi_{\mathbf{8}}^{SU(3)} + \chi_{\mathbf{3}}^{SU(3)}$ | 3 |
| $SU(3)$ | $\chi_{\mathbf{8}}^{SU(3)} + \chi_{\mathbf{8}}^{SU(3)}$ | 8 |
| $SU(3)$ | $\chi_{\mathbf{10}}^{SU(3)}$ | 2 |
| $SU(3) \times U(1)$ | $\chi_{\mathbf{3}}^{SU(3)} t + \chi_{\mathbf{3}}^{SU(3)} \frac{1}{t}$ | 0 |
| $SU(3) \times U(1)$ | $\chi_{\mathbf{3}}^{SU(3)} t + \chi_{\mathbf{\bar{3}}}^{SU(3)} \frac{1}{t}$ | 1 |
| $SU(3) \times U(1)$ | $\chi_{\mathbf{6}}^{SU(3)} t + \chi_{\mathbf{\bar{6}}}^{SU(3)} \frac{1}{t}$ | 3 |
| $SU(3) \times U(1)$ | $\chi_{\mathbf{8}}^{SU(3)} t + \frac{1}{t}$ | 2 |
| $SU(3) \times U(1)$ | $\chi_{\mathbf{8}}^{SU(3)} t + \chi_{\mathbf{3}}^{SU(3)} \frac{1}{t}$ | 1 |

Table C.3: Dimensions of the manifolds spanned by different representations of $SU(4)$ and $SU(4) \times U(1)$

| Group | Representation | Dimension |
|---|---|---|
| $SU(4)$ | $\chi_{\mathbf{4}}^{SU(4)}$ | 0 |
| $SU(4)$ | $\chi_{\mathbf{4}}^{SU(4)} + \chi_{\mathbf{4}}^{SU(4)}$ | 1 |
| $SU(4)$ | $\chi_{\mathbf{4}}^{SU(4)} + \chi_{\bar{\mathbf{4}}}^{SU(4)}$ | 1 |
| $SU(4)$ | $\chi_{\mathbf{6}}^{SU(4)}$ | 1 |
| $SU(4)$ | $\chi_{\mathbf{6}}^{SU(4)} + \chi_{\mathbf{6}}^{SU(4)}$ | 3 |
| $SU(4)$ | $\chi_{\mathbf{10}}^{SU(4)}$ | 1 |
| $SU(4)$ | $\chi_{\mathbf{10}}^{SU(4)} + \chi_{\mathbf{10}}^{SU(4)}$ | 5 |
| $SU(3)$ | $\chi_{\mathbf{10}}^{SU(4)} + \chi_{\overline{\mathbf{10}}}^{SU(4)}$ | 5 |
| $SU(4)$ | $\chi_{\mathbf{15}}^{SU(4)}$ | 6 |
| $SU(4) \times U(1)$ | $\chi_{\mathbf{4}}^{SU(4)} t + \chi_{\bar{\mathbf{4}}}^{SU(4)} \frac{1}{t}$ | 1 |
| $SU(4) \times U(1)$ | $\chi_{\mathbf{15}}^{SU(4)} t + \frac{1}{t}$ | 6 |

Table C.4: Dimensions of the manifolds spanned by different representations of $SO(7)$ and $SO(7) \times U(1)$

| Group | Representation | Dimension |
|---|---|---|
| $SO(7)$ | $\chi_{\mathbf{7}}^{SO(7)}$ | 1 |
| $SO(7)$ | $\chi_{\mathbf{7}}^{SO(7)} + \chi_{\mathbf{7}}^{SO(7)}$ | 3 |
| $SO(7)$ | $\chi_{\mathbf{21}}^{SO(7)}$ | 3 |
| $SO(7)$ | $\chi_{\mathbf{27}}^{SO(7)}$ | 6 |
| $SO(7) \times U(1)$ | $\chi_{\mathbf{7}}^{SO(7)} t + \chi_{\mathbf{7}}^{SO(7)} \frac{1}{t}$ | 2 |
| $SO(7) \times U(1)$ | $\chi_{\mathbf{21}}^{SO(7)} t + \frac{1}{t}$ | 3 |
| $SO(7) \times U(1)$ | $\chi_{\mathbf{27}}^{SO(7)} t + \frac{1}{t}$ | 6 |

Table C.5: Dimensions of the manifolds spanned by different representations of $USp(4)$ and $USp(4) \times U(1)$

| Group | Representation | Dimension |
|---|---|---|
| $USp(4)$ | $\chi_{\mathbf{4}}^{USp(4)}$ | 0 |
| $USp(4)$ | $\chi_{\mathbf{4}}^{USp(4)} + \chi_{\mathbf{4}}^{USp(4)}$ | 1 |
| $USp(4)$ | $\chi_{\mathbf{5}}^{USp(4)} + \chi_{\mathbf{5}}^{USp(4)}$ | 3 |
| $USp(4)$ | $\chi_{\mathbf{5}}^{USp(4)}$ | 1 |
| $USp(4)$ | $\chi_{\mathbf{10}}^{USp(4)}$ | 2 |
| $USp(4)$ | $\chi_{\mathbf{14}}^{USp(4)}$ | 4 |
| $USp(4)$ | $\chi_{\mathbf{16}}^{USp(4)}$ | 6 |
| $USp(4)$ | $\chi_{\mathbf{20}}^{USp(4)}$ | 10 |
| $USp(4) \times U(1)$ | $\chi_{\mathbf{4}}^{USp(4)} t + \chi_{\mathbf{4}}^{USp(4)} \frac{1}{t}$ | 1 |
| $USp(4) \times U(1)$ | $\chi_{\mathbf{5}}^{USp(4)} t + \frac{1}{t}$ | 1 |
| $USp(4) \times U(1)$ | $\chi_{\mathbf{5}}^{USp(4)} t + \chi_{\mathbf{5}}^{USp(4)} \frac{1}{t}$ | 2 |
| $USp(4) \times U(1)$ | $\chi_{\mathbf{10}}^{USp(4)} t + \frac{1}{t}$ | 2 |

Table C.6: Dimensions of the manifolds spanned by different representations of $USp(6)$ and $USp(6) \times U(1)$

| Group | Representation | Dimension |
|---|---|---|
| $USp(6)$ | $\chi_{\mathbf{6}}^{USp(6)}$ | 0 |
| $USp(6)$ | $\chi_{\mathbf{6}}^{USp(6)} + \chi_{\mathbf{6}}^{USp(6)}$ | 1 |
| $USp(6)$ | $\chi_{\mathbf{14}}^{USp(6)}$ | 2 |
| $USp(6)$ | $\chi_{\mathbf{14}}^{USp(6)} + \chi_{\mathbf{14}}^{USp(6)}$ | 8 |
| $USp(6)$ | $\chi_{\mathbf{14}}^{USp(6)} + \chi_{\mathbf{14}}^{USp(6)} + \chi_{\mathbf{6}}^{USp(6)}$ | 13 |
| $USp(6)$ | $\chi_{\mathbf{14'}}^{USp(6)}$ | 1 |
| $USp(6)$ | $\chi_{\mathbf{14'}}^{USp(6)} + \chi_{\mathbf{14'}}^{USp(6)}$ | 7 |
| $USp(6)$ | $\chi_{\mathbf{21}}^{USp(6)}$ | 3 |
| $USp(6) \times U(1)$ | $\chi_{\mathbf{6}}^{USp(6)} t + \frac{1}{t}$ | 0 |
| $USp(6) \times U(1)$ | $\chi_{\mathbf{6}}^{USp(6)} t + \chi_{\mathbf{6}}^{USp(6)} \frac{1}{t}$ | 1 |
| $USp(6) \times U(1)$ | $\chi_{\mathbf{14}}^{USp(6)} t + \frac{1}{t}$ | 2 |
| $USp(6) \times U(1)$ | $\left(\chi_{\mathbf{14}}^{USp(6)} + \chi_{\mathbf{14}}^{USp(6)}\right) t + \frac{1}{t}$ | 8 |
| $USp(6) \times U(1)$ | $\chi_{\mathbf{14'}}^{USp(6)} t + \frac{1}{t}$ | 1 |

These calculations appear in the accompanying Mathematica notebooks [53]. Discussion of the results:

- We can empirically observe that for large enough generic representations (regardless of the Lie group $G$) the dimension of the manifold DM is given by the formula

$$\text{DM} = \dim(\text{reps}) - \dim(G) = \sum_i \dim(R_i) - \dim(\text{adj}). \quad \text{(C.3)}$$

The formula fails to give the correct result for representations smaller or comparable in size to the adjoint representation of $G$, in case the group $G$ does not have a $U(1)$ factor.

- If the group $G$ does have a $U(1)$ factor, there are additional restrictions to the applicability of the formula. According to the formula, we expect that adding a $U(1)$ factor to the group will decrease the DM by one (because $\dim(G)$ increases by one) if no new representations are added, and not change the DM if an additional singlet is added (that compensates the increase in $\dim(G)$). This indeed happens for large enough representations, where for small representations the DM might be larger or smaller than expected. However, in the following cases even having large representations will not be enough:

- If the representations were not charged at all under $U(1)$, then trivially the DM would not be affected (whereas the formula would predict it to decrease by one).
- If the $U(1)$ charges were not oppositely charged (such that $Q_1 \cdot Q_2 < 0$ in the case of two representations), the DM would always be zero since no $U(1)$ singlets could be formed at all.

- We did not calculate the dimension for the spin representation $\chi_{\mathbf{8}}^{SO(7)}$ since it is given in terms of square roots of the maximal torus variables [65] and the residue calculation fails.

- For large enough representations the calculation failed or became numerically prohibitive.

- Even with all the above caveats we cannot guarantee that the formula will hold for any representation or set of representations that we did not calculate directly, but we are going to assume that it holds for the theories we analyze in the thesis.

[13] F. Benini, Y. Tachikawa and D. Xie, *Mirrors of 3d Sicilian theories*, *Journal of High Energy Physics* (2010) [1007.0992].

[14] N. Festuccia and N. Seiberg, *Rigid supersymmetric theories in curved superspace*, *Journal of High Energy Physics* (2011) [1105.0689].

[15] L. Rastelli and S. S. Razamat, *The supersymmetric index in four dimensions*, *Journal of Physics A: Mathematical and Theoretical* **50** (2017) [1608.02965].

[16] B. Willett, *Localization on three-dimensional manifolds*, *Journal of Physics A: Mathematical and Theoretical* **50** (2017) [1608.02958].

[17] C. Beem and A. Gadde, *The superconformal index of N = 1 class S fixed points*, *Journal of High Energy Physics* (2012) [1212.1467].

[18] S. S. Razamat and G. Zafrir, *Exceptionally simple exceptional models*, *Journal of High Energy Physics* (2016) [1609.02089].

[19] C. Córdova, T. T. Dumitrescu and K. Intriligator, *Deformations of superconformal theories*, *Journal of High Energy Physics* (2016) [1602.01217].

[20] S. S. Razamat, O. Sela and G. Zafrir, *Between Symmetry and Duality in Supersymmetric Quantum Field Theories*, *Physical Review Letters* **120** (2018) [1711.02789].

[21] S. S. Razamat, O. Sela and G. Zafrir, *Curious patterns of IR symmetry enhancement*, *Journal of High Energy Physics* (2018) [1809.00541].

[22] S. J. Gates, M. T. Grisaru, M. Rocek and W. Siegel, *Superspace or one thousand and one lessons in Supersymmetry*. Addison-Wesley, 1983.

[23] O. Aharony, A. Hanany, K. Intriligator, N. Seiberg and M. J. Strassler, *Aspects of N=2 Supersymmetric Gauge Theories in Three Dimensions*, *Nuclear Physics B* (1997) [9703110].

[24] I. Yaakov, "Localization of Gauge Theories on the Three-Sphere." Ph.D thesis, Caltech, http://thesis.library.caltech.edu/7116/, 2012.

[25] B. Willett, "Localization and dualities in three-dimensional superconformal field theories." Ph.D thesis, Caltech, http://thesis.library.caltech.edu/7111/, 2012.

[26] B. de Wit, "Supergravity." Lectures given at Les Houches Summerschool. Notes: https://arxiv.org/abs/hep-th/0212245, 2002.

[27] J. Strathdee, *Extended Poincare Supersymmetry*, *International Journal of Modern Physics A* (1987) .